\documentclass{jfm}
\usepackage{graphicx,floatrow}
\usepackage{dcolumn}
\usepackage{bm}
\usepackage{color}

\usepackage{amsmath}

\newcommand{\beq}{\begin{equation}}
\newcommand{\eeq}{  \end{equation}}
\newcommand{\beqa}{\begin{eqnarray}}
\newcommand{\eeqa}{  \end{eqnarray}}
\newcommand{\beas}{\begin{eqnarray*}}
\newcommand{\eeas}{\end{eqnarray*}}

\newcommand{\rf}[1]{(\ref{#1})}

\newcommand{\mR}{\mathbb{R}\,}

\newcommand{\lb}{\label}
\newcommand{\go}{\rightarrow}

\def\strutdepth{\dp\strutbox}
\def\nw#1{\strut\vadjust{\kern-\strutdepth\vtop to0pt{\vss\hbox to\hsize
{\hskip\hsize\hskip5pt$\leftarrow$\hss\strut}}}{\em #1}}

\begin{document}

\title{Thermal rupture of a free liquid sheet}

\author{G. Kitavtsev$^1$ \corresp{\email{georgy.kitavtsev@bristol.ac.uk}}, M. Fontelos$^2$ and J. Eggers$^1$}

\affiliation{
$^1$School of Mathematics, 
University of Bristol, University Walk,
Bristol BS8 1TW, United Kingdom  \\
$^2$Instituto de Ciencias Matem\'aticas,
 (ICMAT, CSIC-UAM-UCM-UC3M), C/ Serrano 123, 28006 
Madrid, Spain.
}

\maketitle

\begin{abstract}
We consider a free liquid sheet, taking into account the dependence of
surface tension on temperature, or concentration of some pollutant.
The sheet dynamics are described within a long-wavelength description.
In the presence of viscosity, local thinning of the
sheet is driven by a strong temperature gradient across the pinch region,
resembling a shock. As a result, for long times the sheet thins exponentially,
leading to breakup. We describe the quasi one-dimensional thickness,
velocity, and temperature
profiles in the pinch region in terms of similarity solutions, which
posses a universal structure. Our analytical description agrees quantitatively
with numerical simulations. 
\end{abstract}

\section{Introduction}
The breakup of liquid sheets plays a crucial role in the generation
of industrial sprays \cite[]{EV08} or natural processes such as sea
spray \cite[]{JWu81}. The
industrial production sprays proceeds typically via the formation of
sheets \cite[]{EV08}, which break up to form ribbons. Ribbons are
susceptible to the Rayleigh-Plateau instability, and quickly break
up into drops. In nature, sheets are often formed by bubbles rising to
the surface of a pool \cite[]{B-SB93,LV17}. Once broken, the sheet decays 
into a mist of droplets \cite[]{LV17,FRVASGTS14}, and collapse of the
void left by the bubble produces a jet \cite[]{DPJZ02}. 

It is therefore of crucial importance to understand the mechanisms
leading to the breakup of sheets. In contrast to jets and liquid threads,
there is no obvious {\it linear} mechanism for sheet breakup, unless
there is strong shear, and the mechanism is that of the Kelvin-Helmholtz
instability \cite[]{TSLMS11}. As a result, authors have invoked the
presence of attractive van-der-Waals forces \cite[]{V66} to explain 
spontaneous rupture \cite[]{TTTE12}. However, the mean sheet thickness 
near the point of breakup is often found to be several microns, while
van-der-Waals forces only have a range of nanometers, and cannot play
a significant role except perhaps for the very last stages of breakup. 

Instead, it has been suggested \cite[]{TB05,BT13,LV17,NV17} that gradients
of temperature could promote breakup, because they produce Marangoni
forces \cite[]{CM09}, which lead to flow. 
This cannot be a linear mechanism, since for reasons of thermodynamic
stability Marangoni flow will always act to reduce gradients; molecular
diffusion will also alleviate (temperature) gradients. Finally, the
extensional flow expected near a potential pinch point will stretch
the fluid particles, once more tending to reduce gradients. It is
therefore surprising that temperature gradients can promote breakup,
and if this is the case, the mechanism must be inherently non-linear. 

In the absence of viscosity, it was found numerically that sheets can break up
in finite time \cite[]{M76,PS98}, if there is a sufficiently strong
initial flow inside the sheet. This was confirmed analytically by
\cite[]{BT07}, who found a similarity solution leading to finite-time
breakup. Their solution is slender, so a long-wavelength approximation can be
used, and the final stages of breakup are described by a local mechanism.
However it is found numerically \cite[]{BT13} and supported by
theoretical arguments \cite[]{EF_book}, that an arbitrary small amount
of viscosity inhibits this singularity, and the sheet returns ultimately
to its original equilibrium thickness. Scaling arguments suggest that the
minimum thickness reached is in the order of the viscous length scale
$\ell_{\nu} = \nu^2\rho/\gamma$ \cite[]{ED94}, which even for a low viscosity
liquid such as water only reaches about 10 nm.

\cite{BT13} have thus asked the question whether in the non-linear regime,
temperature gradients could remain effective in driving the sheet toward
vanishing thickness. If there is no viscosity, yet temperature (and thus
surface tension) gradients are taken into account, the \cite{BT07}
singularity is recovered, and surface tension gradients play a subdominant
role. This is consistent with the above argument that a pinching solution
will only stretch, and thus alleviate, thermal gradients. However
paradoxically, numerical evidence suggests \cite[]{BT13} that if both
finite viscosity and surface tension gradients are taken into account,
breakup can occur, by a mechanism different from those considered
previously. However,  \cite{BT13} were unable to find a consistent
similarity description, and numerical evidence is inconclusive as to 
whether there is a finite time or infinite time singularity. 
Let us also mention a recent study \cite[]{NV17} of the initial
stages of sheet rupture from both an experimental and
an analytical point of view. In particular, the authors 
provide an explanation for the formation of a sharp temperature jump
within the thin pinch-off region, which has been observed by~\cite{BT13}  
to persist during the later self-similar evolution of the sheet.

\begin{figure}
\centering
\includegraphics*[width=0.48\textwidth]{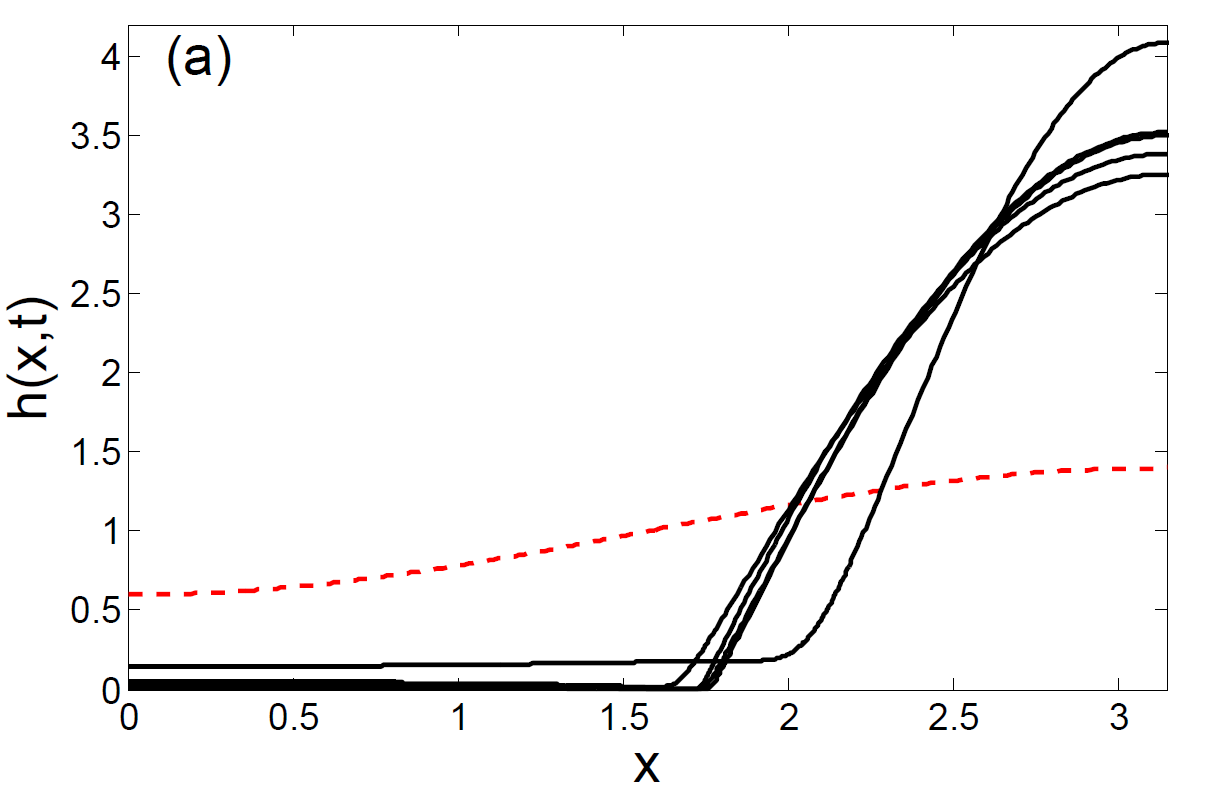}
\includegraphics*[width=0.47\textwidth]{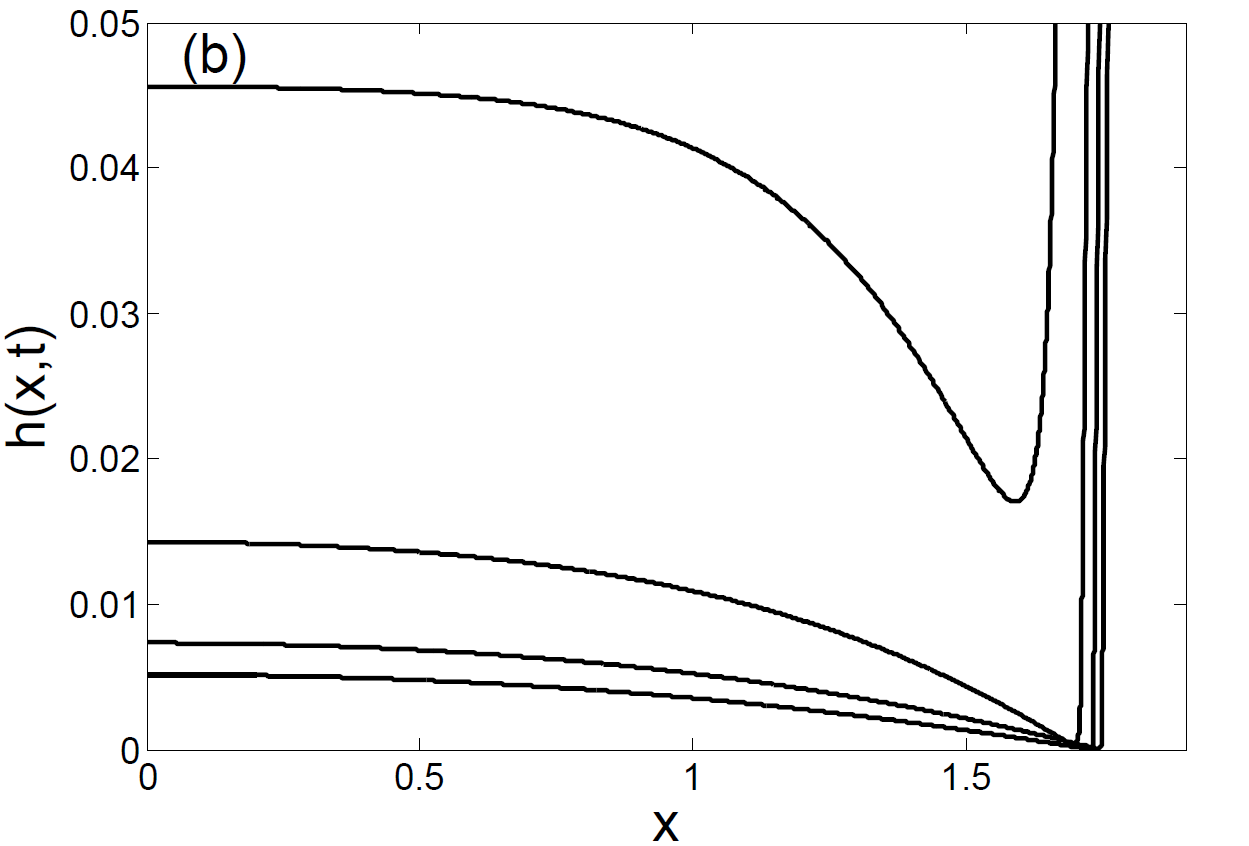}
\includegraphics*[width=0.47\textwidth]{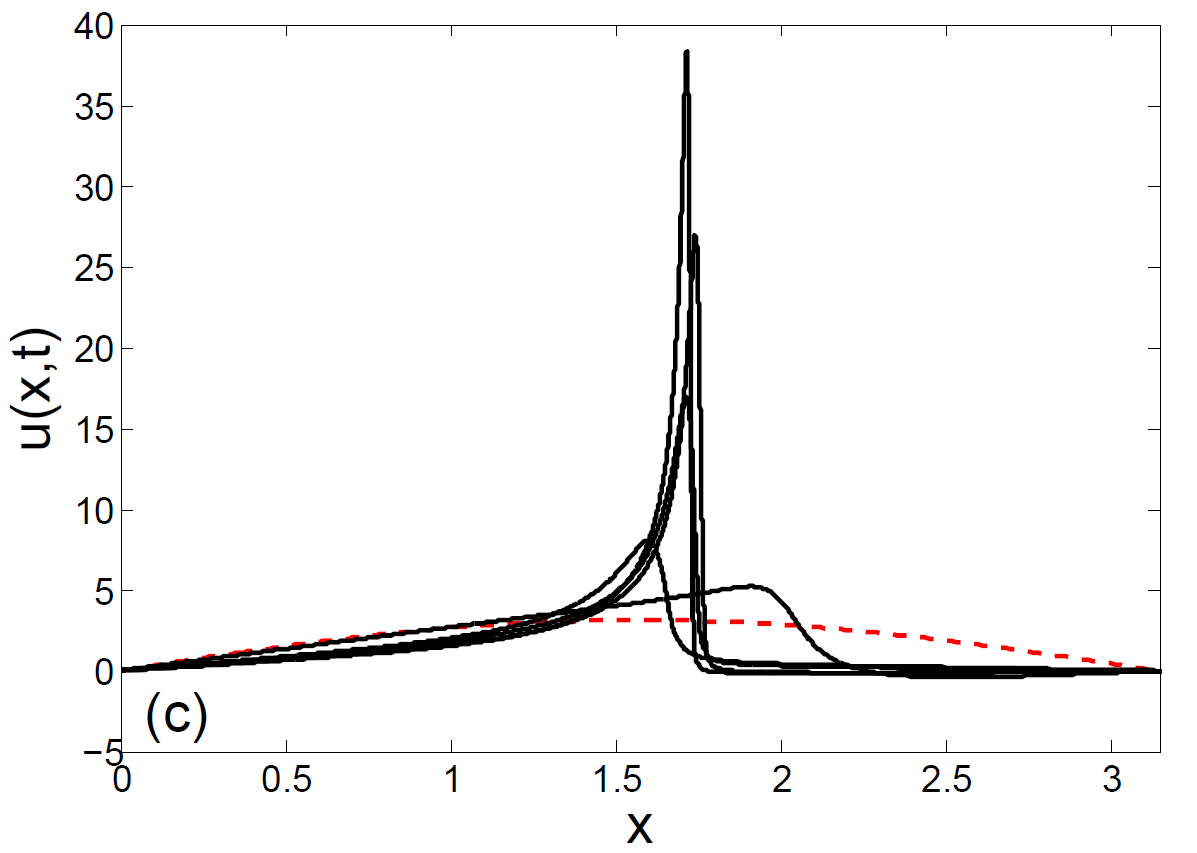}
\includegraphics*[width=0.47\textwidth]{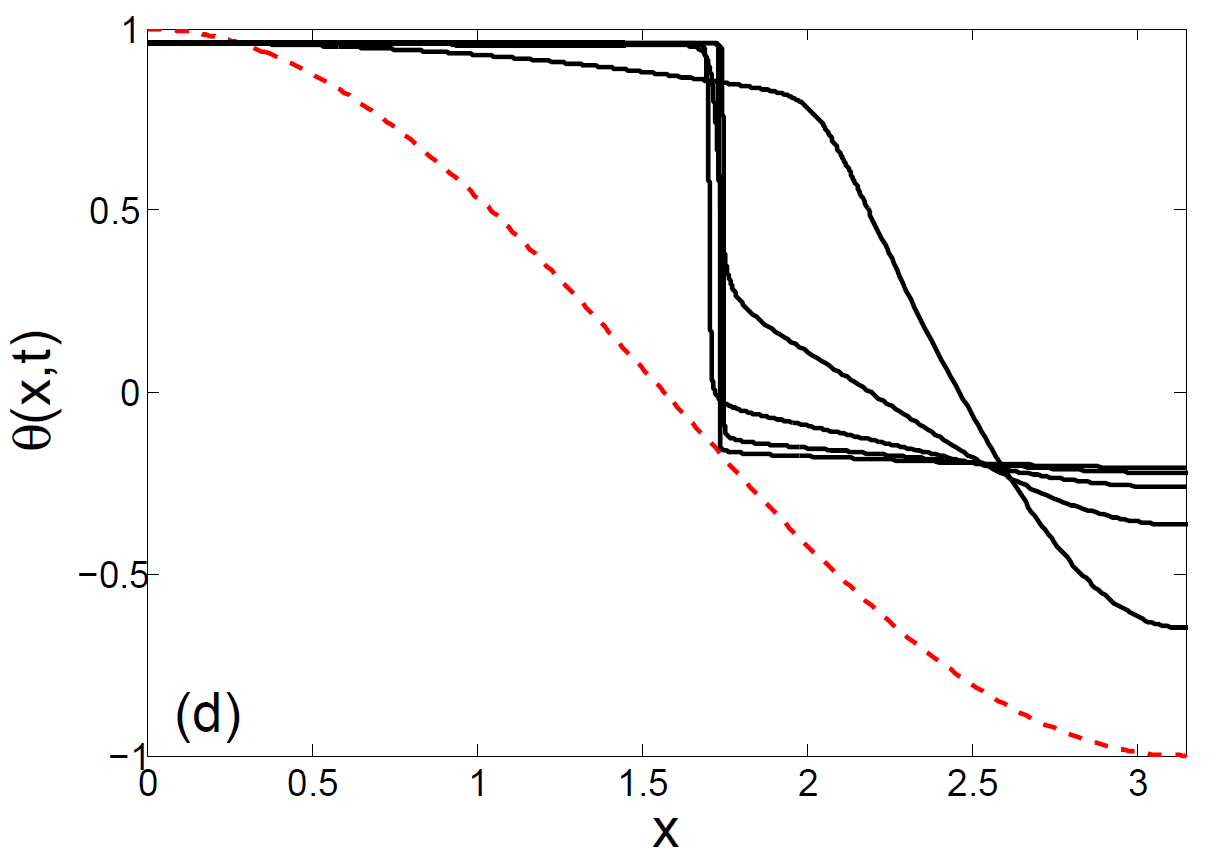}
\caption{Rupture of a viscous liquid sheet as described by \rf{SSM},
starting from initial conditions $h(x,0)=1-0.4\cos(x)$, $u(x,0)=\pi\sin(x)$,
and $\theta(x,0)=\cos(x)$, a particular case of those used in
\cite{BT13} (red dashed curves), with parameters ${\cal O}={\cal D}=1/4$
and ${\cal M}=10$. Shown are six snapshots, taken at times
$t_0=0,\,t_1=2.247,\,t_2=4.032,\,t_3=5.138,\,t_4=5.987,\,t_5=6.800$, of 
the height profile $h(x,t)$ (a), with a zoom of the pinch region shown
in (b). The velocity $v(x,t)$ and temperature $\theta(x,t)$ profiles are shown in (c) and (d), respectively.}
\label{pinch_ex}
\end{figure}

In this paper, we address the late stages of pinch-off in the presence
of both finite viscosity and surface tension gradients. For simplicity,
here we only consider variations of the temperature. These are the equations
for a surfactant in the limit of high solubility \cite[]{JG93,Ma02}.
In the next section,
we describe the equations coming from a long-wavelength assumption:
the sheet thickness is much smaller than a typical variation in the
lateral direction. This description is one-dimensional, in that
gradients in only one direction along the sheet are considered important.
We also describe a finite differences numerical code and show some typical
solutions leading to rupture, an example of which is shown in
Fig.~\ref{pinch_ex}. Starting from smooth initial profiles for the
sheet thickness $h(x,t)$, velocity $u(x,t)$, and temperature
$\theta(x,t)$, the shape of the sheet evolves toward a thin film
on the left, connected to a macroscopic droplet right, see
Fig.~\ref{pinch_ex} (a). A zoom of the pinch region (Fig.~\ref{pinch_ex} (b))
shows that the sheet thickness goes to zero in a localized fashion
near the point where the sheet and the drop meet. In the same region, 
the velocity has a sharp and increasing maximum  (Fig.~\ref{pinch_ex} (c)), while the temperature  
develops an increasingly sharp jump  (Fig.~\ref{pinch_ex} (d)). 

In the third section, we construct an analytical solution 
in which the sheet thickness goes to zero exponentially. The macroscopic
outer part consists of an exponentially thinning film, on one side,
and a static ``bubble'' on the other. Over both parts of the outer
solution the temperature is approximately constant but different, with
a strong gradient between the two regions. The pinch region connecting
the two parts is described by two different similarity solutions, which
hold in two different regions, with two different sets of scaling exponents.
Matching all regions together, we are able to describe pinch-off in terms
of a single free parameter, which is the position of the pinch point.
All other parameters are found in terms of the initial conditions, or can
be absorbed into a shift in time. The results agree very well with numerical
simulations of the long-wavelength equations. We show for the first time,
both theoretically and numerically (see Fig.~\ref{pinch_ss_law}), 
that the minimum of the sheet thickness $h(x,t)$ decreases exponentially,
at a rate we calculate. In a final section, we discuss
our results and give perspectives. The Appendix presents a detailed
analysis of the leading order equation arising in the exponentially
thinning film region and contains a complete list of its possible solutions.

\section{Long-wavelength equations and simulation}
We consider the motion of a free liquid sheet, whose plane of symmetry
has been fixed in the $z$-plane. We expect that generically the sheet 
breaks up along a line, so in describing this singularity, we can
assume that fields only depend on the coordinate $x$ perpendicular to
this line. Thus the shape of the sheet is described uniquely by the
half-thickness $h(x,t)$. We assume that the surface tension depends
linearly on temperature $\theta$ \cite[]{CM09} according to
\beq
\gamma = \gamma_0 - k\theta,
\label{gamma}
\eeq
which is a good approximation away from any critical point
\cite[]{Rowlinson}. We will assume that $k >0$ as is the case for
most systems, but the opposite sign will simply reverse the flow
of heat. The average velocity in the sheet is $u(x,t)$, and the
temperature $\theta(x,t)$, which is allowed to diffuse.

Then in the limit of slender sheets \cite[]{BT13}, the
dimensionless form of the equations is 
\begin{subequations}
\label{SSM}
\begin{align}
h_t &= -\left(h u\right)_x
\label{SSM2}\\
u_t+uu_x&= h_{xxx}+4{\cal O}\frac{(h u_x)_x}{h}
- {\cal M}\frac{\theta_x}{h}
\label{SSM1}\\
\theta_t+u\theta_x&={\cal D}(h\theta_x)_x,
\label{SSM3}
\end{align}
\end{subequations}
where subscripts denote differentiation with respect to the
variable. As a length scale, we have chosen $L_0=L/\pi$, where $L$ is the
width of the computational domain, and
$\tau = \sqrt{L_0^3\rho/\gamma_0}$ is the time scale; $\rho$ is the
fluid density. As a unit of temperature we take the initial
temperature difference $\Delta$ across the system.

Then \eqref{SSM2} describes mass conservation, and \eqref{SSM1} is
the momentum balance across the sheet. Inertial forces on the
left are balanced by surface tension, viscous stresses, and Marangoni
forces on the right, respectively. The size of the kinematic viscosity
$\nu$ is measured by the Ohnesorge number
${\cal O} = \nu\sqrt{\rho/(L_0\gamma_0)}$, and the Marangoni number is
defined by ${\cal M} = k\Delta\gamma_0$. We assume that the variation
of the surface tension is small, so we can take it as a constant,
except in the Marangoni term.
The last equation \eqref{SSM3} describes the diffusion of temperature
through the sheet, and ${\cal D} = \gamma \sqrt{\rho/(L_0\gamma_0)}$,
where $\kappa$ is the thermal diffusion coefficient;
${\cal P} = {\cal O}/{\cal D}$ is known as the Prandtl number. 

For simplicity, we consider solutions to \eqref{SSM} in a fixed domain
$[0,\pi]$ (after non-dimensionalization), and assume that
\beq
u = h_x=\theta_x = 0\; \mbox{for}\; x=0,\pi,
\lb{bc}
\eeq
i.e. no-flux boundary conditions for the velocity and free boundary
conditions for the height and the temperature. This choice of boundary
conditions is motivated by the fact that for symmetric initial data,
for example those of Fig.~\ref{pinch_ex}, they result in solutions which
can be extended to periodic solutions of period $2\pi$. The conditions
\rf{bc} also imply that there is no mass or heat flux out of the system,
and thus 
\beq
M = \int_0^\pi h(x,t)\,dx, \quad Q = \int_0^\pi \theta(x,t)h(x,t)\,dx
\lb{CLs}
\eeq
are conserved quantities, set by the initial conditions. We do not expect
our choice of boundary conditions to have an effect on the structure of
the singularity. The outer film and bubble regions will still be described
by the same leading order ODEs, but their solutions may be selected
by the particular boundary and initial data. However, with the particular
choice of
boundary conditions \rf{bc} we are able to determine the structure of the
singularity largely in terms of the two quantities $M$ and $Q$ alone.

Our main focus will be on pinch-off singularities for which
$h\rightarrow 0$ at some point $x_0$ in space. To summarize what is known
or widely accepted about pinch-off singularities of the system
\eqref{SSM}, and as stated in the Introduction, for ${\cal O}=0$ 
finite-time pinch-off can occur for suitable initial conditions.
The neighborhood of the pinch point is described by the similarity
solution of \cite{BT07} for any value of ${\cal M}$, and Marangoni
forces are subdominant. If on the other hand ${\cal O}$ is finite
and ${\cal M}=0$, breakup can never occur \cite[]{EF_book} and instead
the sheet will eventually relax to a uniform state $h(x) = M$. 
The present paper deals with the case that both ${\cal M}$ and
${\cal O}$ are nonzero, for which we find a local pinch solution for
which the thickness goes to zero exponentially in time
(a typical example being presented in Fig. \ref{pinch_ex}).

To solve the system \rf{SSM} we extended the finite-difference schemes
developed previously for the modeling of finite
time rupture under the presence of van der Waals forces
by~\cite{Peschka08,PMN10} and of coarsening dynamics of droplets in free
liquid films by~\cite{KW10}. We incorporated the temperature equation
\rf{SSM3} along with the Marangoni term -${\cal M}\theta_x/h$,
coupled with the boundary conditions \rf{bc}. The resulting fully implicit
finite-difference scheme is solved on a general nonuniform mesh in space,
with adaptive time step. At every time step the nonlinear system
of algebraic equations is solved using Newton's method. In order to resolve
the solution close to the rupture point we applied the algorithm of
\cite{Peschka08} for dynamical  grid re-meshing to concentrate points
near where the film thickness is the smallest.

\section{Self-similar pinch-off solutions}
\begin{figure}
\includegraphics*[width=0.8\textwidth]{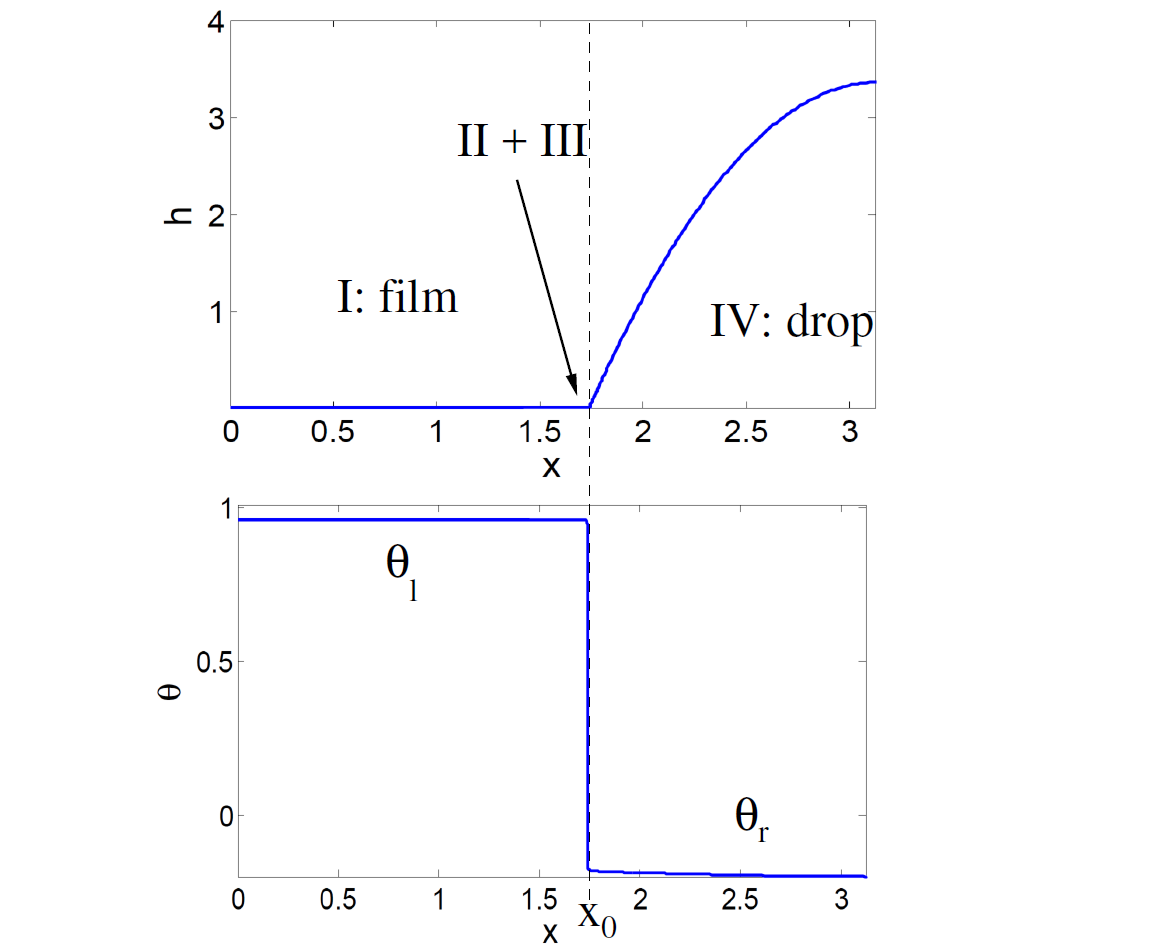}
\caption{Typical height (top) and temperature (bottom)
profiles near pinch-off. On the left, a film (region I) which thins
exponentially, on the right a drop in static equilibrium (region IV).
These two outer regions
are joined together at $x_0$, where pinch-off ultimately occurs. 
The temperature inside the film has a nearly constant value $\theta_l$,
in the drop a constant value $\theta_r$, with a sudden drop in the pinch
region. The inner region at the juncture between film and drop has to
split into two sub regions (II and III), which have different scalings. 
}
\label{overview}
\end{figure}

We begin with an overview of the structure of the solution in the
asymptotic region we hope to describe, see Fig.~\ref{overview}. The
outer solution, observed on a macroscopic scale, is split between
a thin film region I on the left, and a drop region IV on the right;
the two are joined together at the pinch point $x_0$. The film thins
exponentially in time, while the drop is in static equilibrium, and
has a stationary profile. The temperature is almost constant in the
two regions, with a sudden jump near the pinch point. Since the surface
tension is lower on the left (higher temperature), this drives a Marangoni
flow from the film into the drop, which is responsible for the thinning of
the film.

The crucial question is how this strong temperature gradient is
maintained, and what stabilizes the sudden jump in temperature.
To understand this, one must study the inner region joining the two
outer solutions, whose width will turn out to be of the same order as
the film thickness, and which is therefore not resolved in
Fig.~\ref{overview}. It turns out that in order to achieve a matching between
regions I and IV, one must subdivide the inner region into two sub-regions, 
characterized by similarity solutions with different scalings.
The first one, region II, we call the ``pinch region'', because the film
thickness has its minimum there and it is where pinch-off
ultimately occurs. This region is characterized by a balance of inertia,
viscosity, and Marangoni forces. However this does not match the drop
region, where surface tension alone is important. This necessitates
another region III, the ``transition region'', where only surface tension and
viscosity are important. 

The fundamental insight which determines the structure of both
similarity solutions is that the flux of liquid across the inner
region is set by the flux out of the thin film region, which is
set on a macroscopic scale. Thus the flux $j = hu$ across inner regions
II and III must be a spatial constant (but does depend on time). We will
see that this constraint fixes the scaling exponents, and greatly simplifies
the structure of
the solution. Curiously, a similar structure was found for Hele-Shaw
flow \cite[]{BBDK94} and viscous films in a capillary tube \cite[]{LE17}. 

We now present all asymptotic regions systematically, and discuss matching
between them. 

\subsection{I: thin film region}
\label{sub:thin}
The width of this region is of order one, yet the thickness
$\tau(t)$ of the film shrinks to zero, so we use the ansatz
\beq
h(x,t)=\tau(t)h_f(x), \quad
u(x,t)=u_f(x), \quad
\theta(x,t)=\theta_l,
\lb{expUTF}
\eeq
where the temperature is assumed constant, in accordance with
our earlier observations. Inserting \eqref{expUTF} into the equations
of motion, \eqref{SSM2} yields
\beq
\dot{\tau}h_f = -\tau\left(h_f u_f\right)',
\label{mass_f}
\eeq
and at leading order $\tau^0$, \eqref{SSM1} results in 
\beq
u_f u_f' = 4{\cal O}\frac{(h_f u_f')'}{h_f}, 
\label{mom_f}
\eeq
where the surface tension term is of order $\tau$, and thus drops
out in the limit $\tau\rightarrow 0$. Here and in the remainder
of the paper, a dot denotes a derivative with respect to time, a
prime with respect to the spatial variable. Note that \eqref{mom_f}
represents a balance between inertia and viscosity, while surface tension
and Marangoni forces drop out. Dividing \eqref{mass_f} by $\tau h_f u_f$
and \eqref{mom_f} by $4{\cal O}u_f'$, the term $h_f'/h_f$ can be eliminated between
the two equations, and one obtains an equation for the velocity $u_f$ alone:
\beq
\bar{u}+\frac{\dot{\tau}}{\tau 4{\cal O}\bar{u}}
= \frac{\bar{u}''}{\bar{u}'} - \frac{\bar{u}'}{\bar{u}}, 
\lb{film_equation}
\eeq
where we have rescaled the velocity according to: $u_f = 4{\cal O} \bar{u}$. 
In \eqref{film_equation} only the second term on the left hand side
depends on time. Therefore, necessarily one has $-\dot{\tau}/\tau=4a{\cal O}$,
where $a>0$ is a constant, which depends on initial conditions, as we will
see. This implies
\beqa
\lb{ODEu_0}
&& \tau(t)=\tau_0\exp\{-4 a {\cal O} t\},\\
&& \bar{u}-\frac{a}{\bar{u}} =
\frac{\bar{u}''}{\bar{u}'}-\frac{\bar{u}'}{\bar{u}},
\lb{ODEu_0b}
\eeqa
where $\tau_0$ is an arbitrary normalization factor, which depends on
the choice of origin for the time coordinate. 

As shown in the Appendix, \rf{ODEu_0b} can be integrated and posses
a one-parameter family of 'blow-up' solutions of the form:
\beq
\bar{u}_A(x)=A + \tan\left[(x-\bar{x})\sqrt{a-A^2}\right]\sqrt{a-A^2},
\quad\text{for}\ A\in(-\sqrt{a},\sqrt{a}).
\lb{gs_film}
\eeq
The boundary conditions $\bar{u}(0)=\bar{u}''(0)=0$, which
follow from \eqref{bc}, together with \eqref{mom_f}, require that
$A = \bar{x} = 0$, and thus 
\beq
u_f = 4{\cal O} \sqrt{a}\tan\left(x\sqrt{a}\right). 
\lb{film_u}
\eeq
The flux $j_f(x) = h_f(x)u_f(x)$ is calculated from \rf{mass_f} as
\beq
\ln j_f =\int\frac{4{\cal O}a}{u_f}dx, 
\lb{flux_film}
\eeq
and then it follows from \eqref{film_u} that 
\beq
j_f(x)=j_0\sin(\sqrt{a}x), \quad
h_f(x)=\frac{j_0}{4{\cal O}\sqrt{a}}\cos(\sqrt{a}x),
\label{Flux}
\eeq
where $j_0$ is a positive constant. It is clear from the first
equation of \rf{expUTF} that by adjusting $\tau_0$, we can make
$h_f(0) = j_0 / 4{\cal O}\sqrt{a}$ attain any value, which means that
$j_0$ can be chosen arbitrarily. In the numerical results reported
below, we will make the particular choice $j_0 = 4{\cal O}\sqrt{a}$.

\begin{figure}
\centering
\includegraphics*[width=0.47\textwidth]{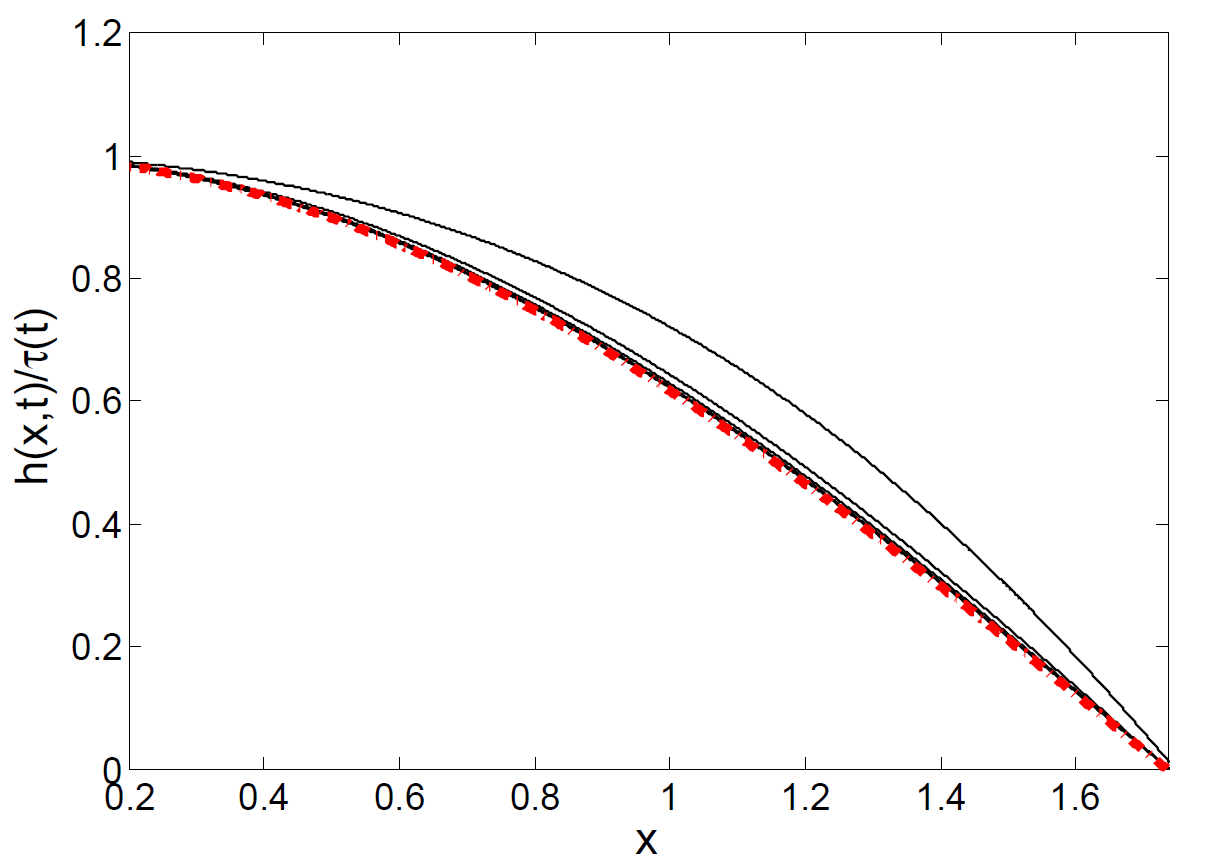}
\includegraphics*[width=0.47\textwidth]{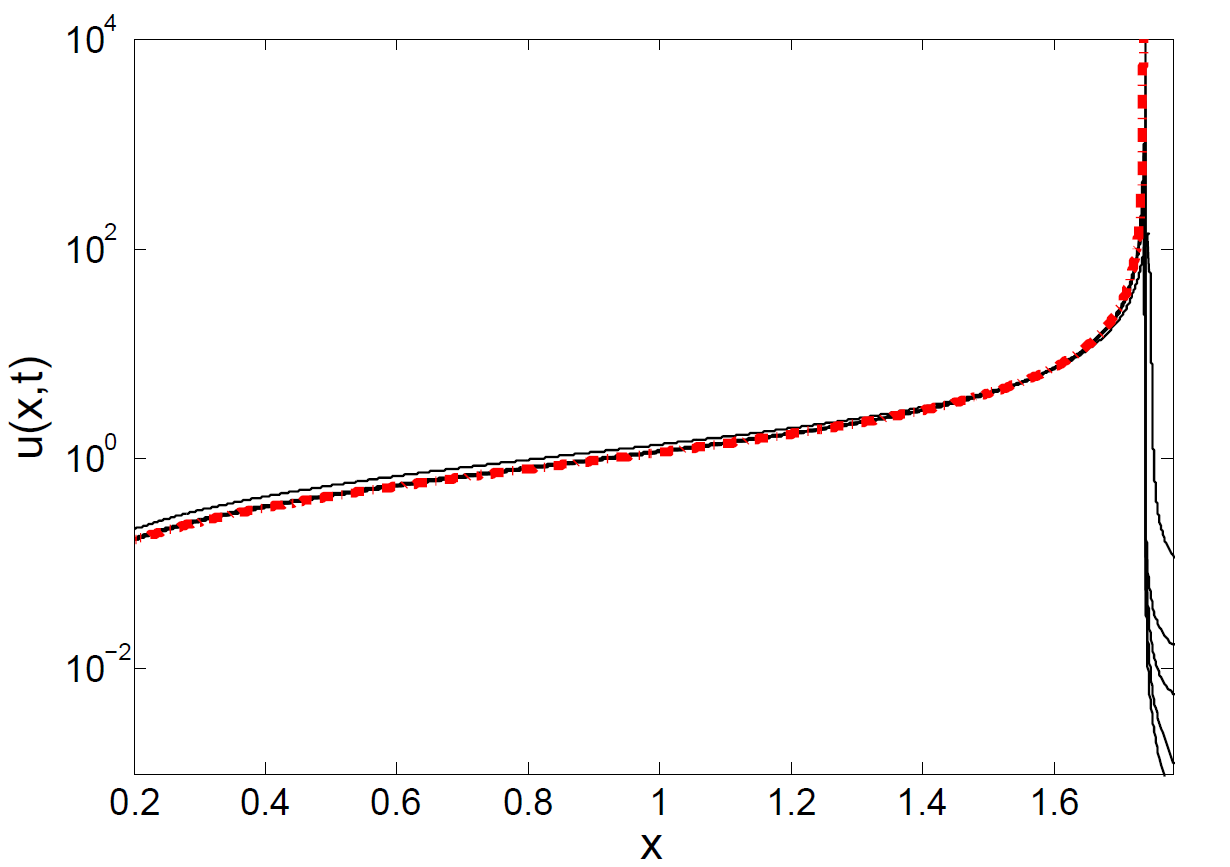}

\caption{Comparison of a numerical solution of \rf{SSM} 
(initial data and parameters as in Fig. \ref{pinch_ex}), with 
the leading order solutions \rf{Flux} and \rf{film_u}
(red dash-dotted curves), in the then film region.
Five snapshots of the height (left, linear plot) and velocity
(right, log-linear plot) solutions to \rf{SSM} are shown at
times $t_1=2.247,\,t_2=4.032,\,t_3=5.138,\,t_4=5.987,\,t_5=6.800$.
The thinning rate $a=0.819$ was calculated from the numerical solution 
using \rf{ODEu_0b}. The pinch point was found to be 
$x_0\approx 1.7355$, in very good agreement with \rf{pp}.}
\label{pinch_film}
\end{figure}
The pinch point $x_0$ is determined by where $h_f$ goes to zero,
which is at
\beq
x_0=\pi/(2\sqrt{a});
\lb{pp}
\eeq
the interval $0\le x \le x_0$ will be referred to as the
``thin film region''. At $x_0$, the flux is $j_f(x_0) = j_0$, which means
that according to \rf{expUTF} the mass flux through the pinch
point and into the drop is $\tau j_0$. In the neighborhood of $x_0$,
the outer (film) profiles are
\beq
h \approx -\frac{j_0\tau(x-x_0)}{4{\cal O}}, \quad
u \approx -\frac{4{\cal O}}{x-x_0}. 
\label{film_out}
\eeq
In Fig.~\ref{pinch_film} we present the leading order film solutions
\rf{film_u} and \rf{Flux} in the thin film region 
(red dash-dotted curves), superimposed with full numerical solutions,
rescaled according to \rf{expUTF}. Even for times of order one,
very convincing convergence toward the asymptotic solutions is found. 

\subsection{IV: drop region}

The total mass in the film region is of order $\tau$, which means
that any change in the volume of the drop region is a subdominant
correction. To leading order, the drop volume is constant and
the drop thus converges toward a static shape, with no flow,
and temperature is constant: $h(x,t) = h_d(x)$
and $\theta(x,t)=\theta_r$, while $u(x,t) = 0$. The leading order solution to \rf{SSM1} in this region
must satisfy $h_d''' =0$, and thus 
\beq
h_d(x)=C_0\left[(\pi-x_0)^2-(x-\pi)^2\right],
\lb{parabola}
\eeq
where $x_0$ is the pinch point as in \eqref{pp}. The constants
$C_0$ and $\theta_r$ are determined uniquely by conservation of
mass and heat \rf{CLs}, which yields
\[
\int_{x_0}^\pi h_d(x)\,dx=M, \quad \int_{x_0}^\pi h_d(x)\theta_r\,dx=Q. 
\]
From this the constants can be computed as 
\beq
C_0=\frac{3M}{2(\pi-x_0)^3},\quad \theta_r=\frac{Q}{M}.
\lb{CLs1}
\eeq
In particular, we have the following expression for the macroscopic
contact angle of the drop:
\beq
h_d^{\prime}(x_0)=\frac{3M}{2(\pi-x_0)^2} \equiv s, 
\lb{ConAng}
\eeq
which will be used later to match to the pinch region.

\subsection{II: pinch region}

Since this solution lives on an exponentially small scale set by the
film thickness $\tau$, we try the similarity solution 
\beq
h(x,t)=\tau^{\alpha_1} H\left(\xi\right),\quad
u(x,t)=\tau^{\alpha_2} U\left(\xi\right),\quad
\theta(x,t)= \tau^{\alpha_3}\Theta\left(\xi\right), \quad
\xi = \frac{x-x_0}{\tau^{\beta}}. 
\lb{SSc}
\eeq
Since the flux through the pinch region is $\tau j_0 = h u$, we must
have $\alpha_1 + \alpha_2 = 1$. We also expect \eqref{SSc} to match
to the linear $h$-profile \eqref{film_out}, which implies that
$\alpha_1 - \beta = 1$. Since the temperature changes over scale of order
unity, we have $\alpha_3 = 0$. Finally, Marangoni forces drive the
pinch-off and thus must come in at leading order near the pinch point.
We expect them to be balanced by viscous forces, which already come
in the thin film region, and thus should also be important on even
smaller scales. Then from a balance of the last two terms of \eqref{SSM1}
we obtain $\alpha_2-2\beta = -\beta-\alpha_1$, and combining all of the
above yields $\beta=1$, $\alpha_1 = 2$, and $\alpha_2 = -1$. Then
the leading force balance in \eqref{SSM1} is at $O(\tau^3)$, and
{\it inertial, surface tension, and Marangoni forces} come in at leading order. 

Thus in the pinch region the similarity solution takes the form
\beq
h(x,t)=\tau^2 H_p\left(\xi\right),\quad
u(x,t)=\tau^{-1} U_p\left(\xi\right),\quad
\theta(x,t)= \Theta_p\left(\xi\right), \quad
\xi = \frac{x-x_0}{\tau},
\lb{SSc_pinch}
\eeq
and the similarity equations are 
\begin{subequations}
\label{SSMr2}
\begin{align}
j_0&=H_pU_p
\label{SSM2r1}\\
U_p U_p'&=4{\cal O}\frac{(H_pU_p')'}{H_p} - {\cal M}\frac{\Theta_p'}{H_p}
\label{SSM1r1}\\
U_p\Theta_p'&={\cal D}\left(\frac{H_p\Theta_p'}{H_p}\right)'.
\label{SSM3r1}
\end{align}
\end{subequations}
\begin{figure}
\centering
\includegraphics*[width=0.6\textwidth]{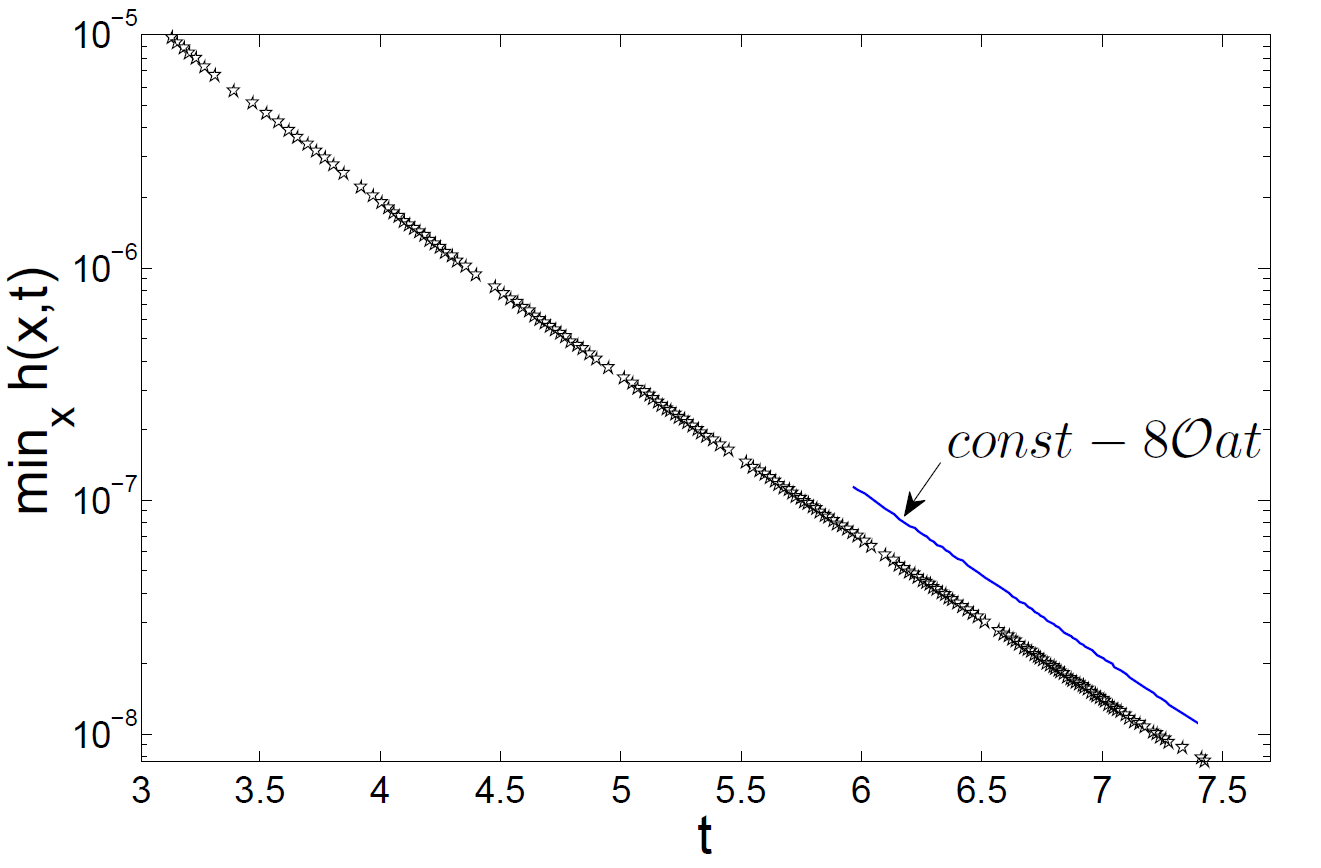}
\caption{Log-linear plot of the minimum height as function of time
(black symbols) for the solution to \rf{SSM}, with parameters and initial
data as in Fig.~\ref{pinch_ex} and Fig.~\ref{pinch_film}. The thinning
rate $a=0.819$ was calculated from the numerical solution using \rf{ODEu_0b}.
The blue solid line is the theoretical prediction \rf{min_thick}.
\label{pinch_ss_law}}
\end{figure}
In particular, using \rf{ODEu_0} the minimum sheet thickness decays
exponentially:
\beq
{\rm min}_x h(x,t) \propto \tau^2 \propto \exp\{-8 a {\cal O} t\},
\label{min_thick}
\eeq
which is confirmed numerically in Fig.~\ref{pinch_ss_law}. 

The flux condition \eqref{SSM2r1} can be used to eliminate $H_p$, and
we obtain the two equations
\beq
U_p' = 4{\cal O}\left(U_p'/U_p\right)' - {\cal M}\Theta_p'/j_0, \quad
\Theta_p'= {\cal D}\left(\Theta_p'/U_p\right)'.
\lb{pinch_combine}
\eeq
Integrating the first equation in \rf{pinch_combine}
one expresses the temperature profile dependence on $U_p$ explicitly as 
\beq
\Theta_p = \frac{j_0}{\cal M}\left(4{\cal O}\frac{U_p'}{U_p}-U_p\right)
+ \theta_l, 
\lb{h_theta}
\eeq
where we have used the boundary
conditions on the similarity profiles, as $\xi\rightarrow -\infty$:
\beq
H_p \approx -\frac{j_0\xi}{4{\cal O}}, \quad
U_p\approx -\frac{4{\cal O}}{\xi},\quad
\Theta_p\approx \theta_l, 
\lb{BCPoff_l}
\eeq
which follow from comparison to \eqref{film_out}.

Next, substitution of \rf{h_theta} into the second (temperature) equation
of \rf{pinch_combine} gives the following ODE for the profile $U_p$:
\beq
4{\cal D}{\cal O} \left(U_p''- U_p'^2/U_p\right) -
(4{\cal O} + {\cal D})U_p'U_p + U_p^3 + C_2U_p^2=0,
\lb{u_e}
\eeq
with $C_2$ being a constant of integration. Evaluating the left-hand side
of \rf{u_e} for $\xi\rightarrow -\infty$, once more using \rf{BCPoff_l},
one concludes that $C_2 = 0$.

\begin{figure}
\centering
\includegraphics*[width=0.6\textwidth]{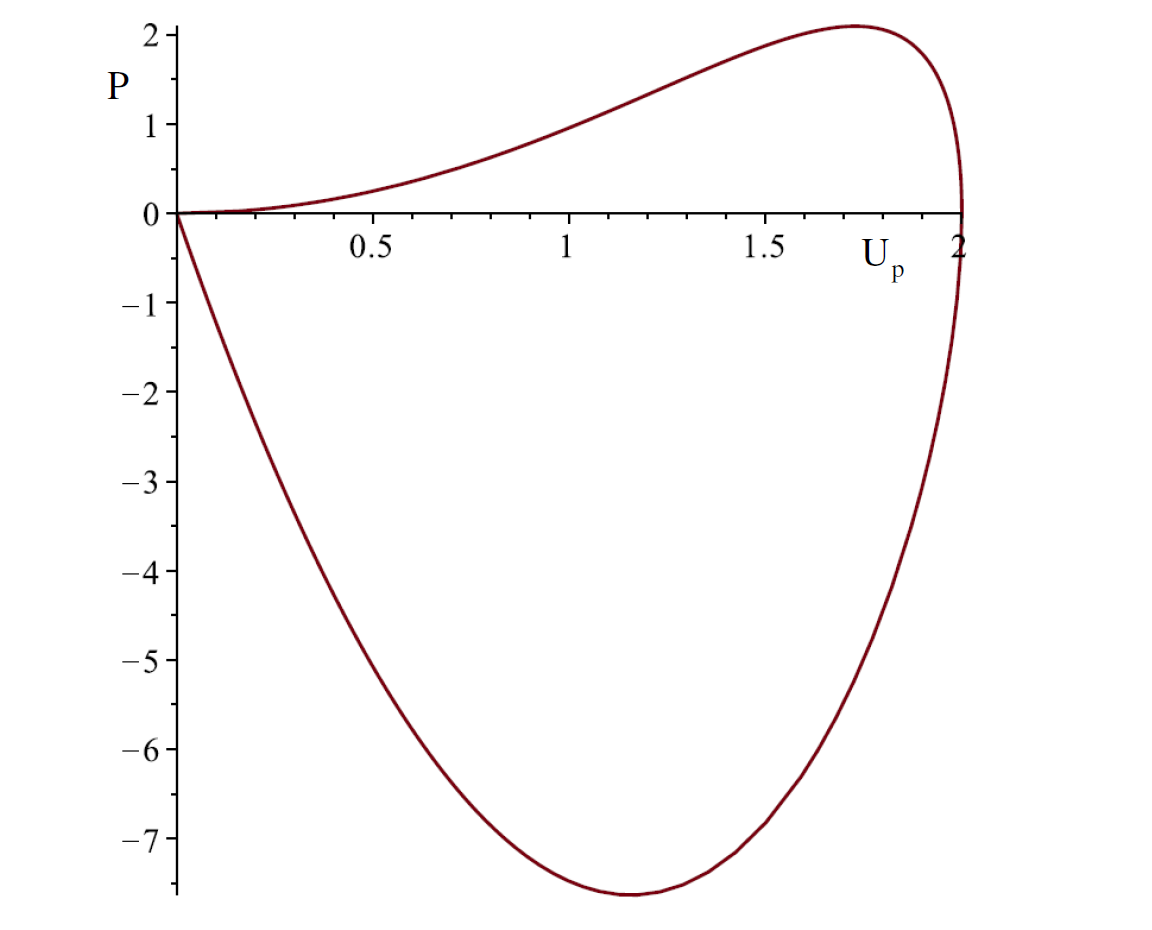}
\caption{The homoclinic orbit defined by \rf{V_U_1}-\rf{V_U_2},
with $C_+=2\cdot4^{4/3}$ and ${\cal P}=1$.}
\label{pinch_phase_plot}
\end{figure}
The second order equation \eqref{u_e} can be turned into a first-order
equation putting $U_p'(\xi) = P(U_p)$, so that $U_p'' = P'P$, and 
\beq
\frac{dP}{dU_p}=\frac{P}{U_p}+\left({\cal D}^{-1}+(4{\cal O})^{-1}\right)U_p-
\frac{U_p^3}{4{\cal D}{\cal O}P} . 
\lb{V_U}
\eeq
The observation that \eqref{V_U} is invariant under 
$U_p\go C_+U_p$ and $P \go C_+^2 P$ suggest the substitution
\beq
P=\frac{w U_p^2}{4\sqrt{{\cal D}{\cal O}}},
\lb{V_U_1}
\eeq
which reduces \rf{V_U} to the separable ODE
\beq
\frac{dU_p}{U_p}=\frac{dw}{4\sqrt{\cal P} + \sqrt{\cal P}^{-1} -4/w - w}. 
\lb{Up_w}
\eeq

From \eqref{BCPoff_l} it follows that $w$ must satisfy the boundary
condition $w \approx \sqrt{\cal P}^{-1}$ for $U_p\rightarrow 0$.
Hence for each ${\cal P}>0$ bounded solutions of  \rf{Up_w} have the form:
\begin{align}
U_p(w)=\left\{ 
\begin{array}{ll}
C_+\left(\sqrt{\cal P}^{-1}-w\right)^{\frac{1}{4{\cal P}-1}}/
\left(4\sqrt{\cal P}-w\right)^{\frac{4{\cal P}}{4{\cal P}-1}}
&\mbox{for } {\cal P}\not=\frac{1}{4},\\
C_+\exp\left\{\frac{2}{w-2}\right\}/(2-w) &\mbox{for } {\cal P}=\frac{1}{4};
\end{array}
\right.
\lb{V_U_2}
\end{align}
they are parameterized by a positive constant $C_+$ and are defined
in the range $w\in(-\infty,\,\sqrt{{\cal P}}^{-1})$. A typical plot of
$P(U_p)$ is shown in Fig.~\ref{pinch_phase_plot}. 
The behavior near the origin on the upper lobe is
$P\approx U_p^2/(4{\cal O})$, and corresponds to $\xi\rightarrow -\infty$,
where $U_p'>0$. This matches the expected asymptotic behavior \eqref{BCPoff_l}.
The lower lobe near the origin, on the other hand,
corresponds to $\xi\rightarrow\infty$, and here
$U\approx -C_+/w$, so that
\[
P\approx -\frac{C_+ U}{4{\cal O}\sqrt{\cal P}^{-1}}, \quad
\xi\rightarrow\infty.
\]

\begin{figure}
\centering
\includegraphics*[width=0.48\textwidth]{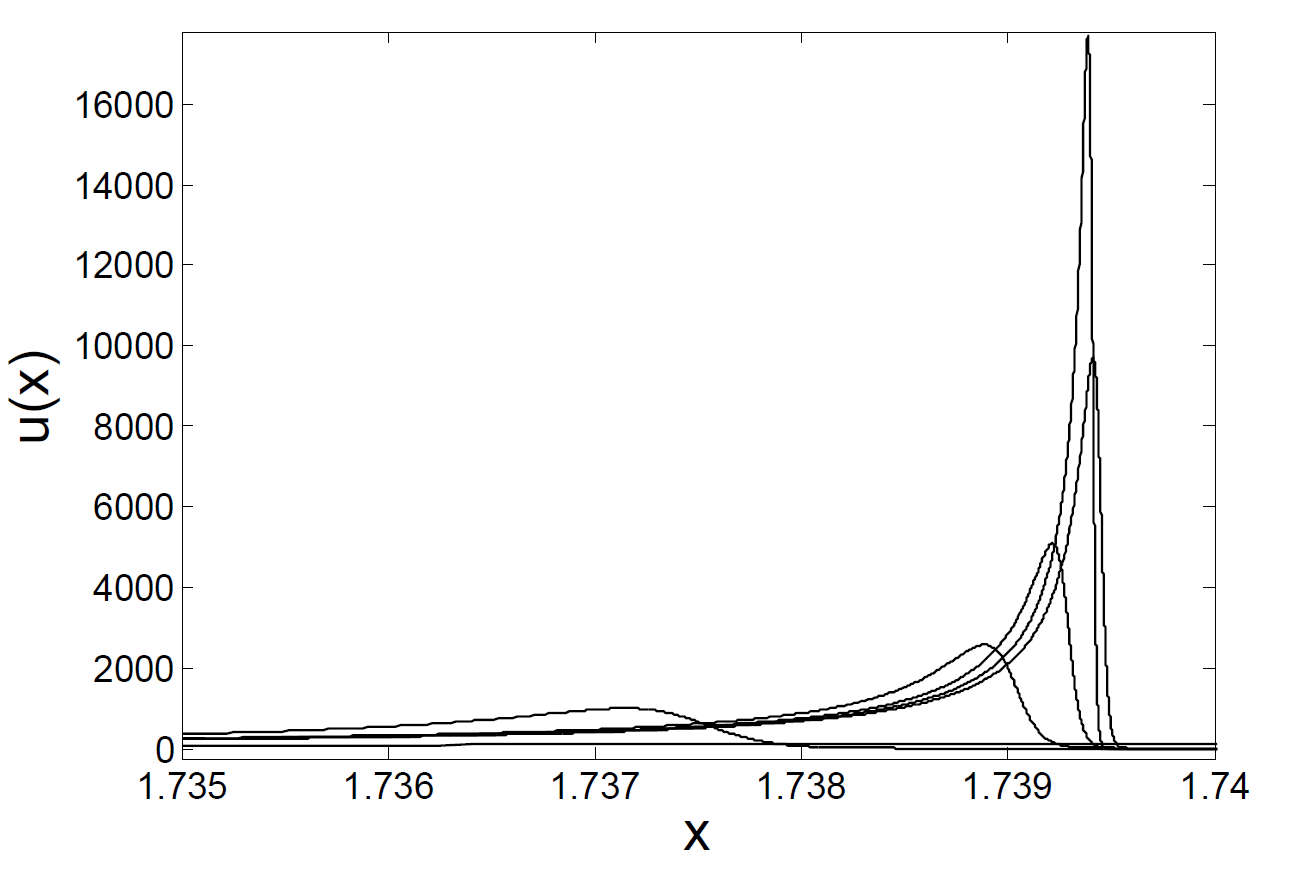}
\includegraphics*[width=0.48\textwidth]{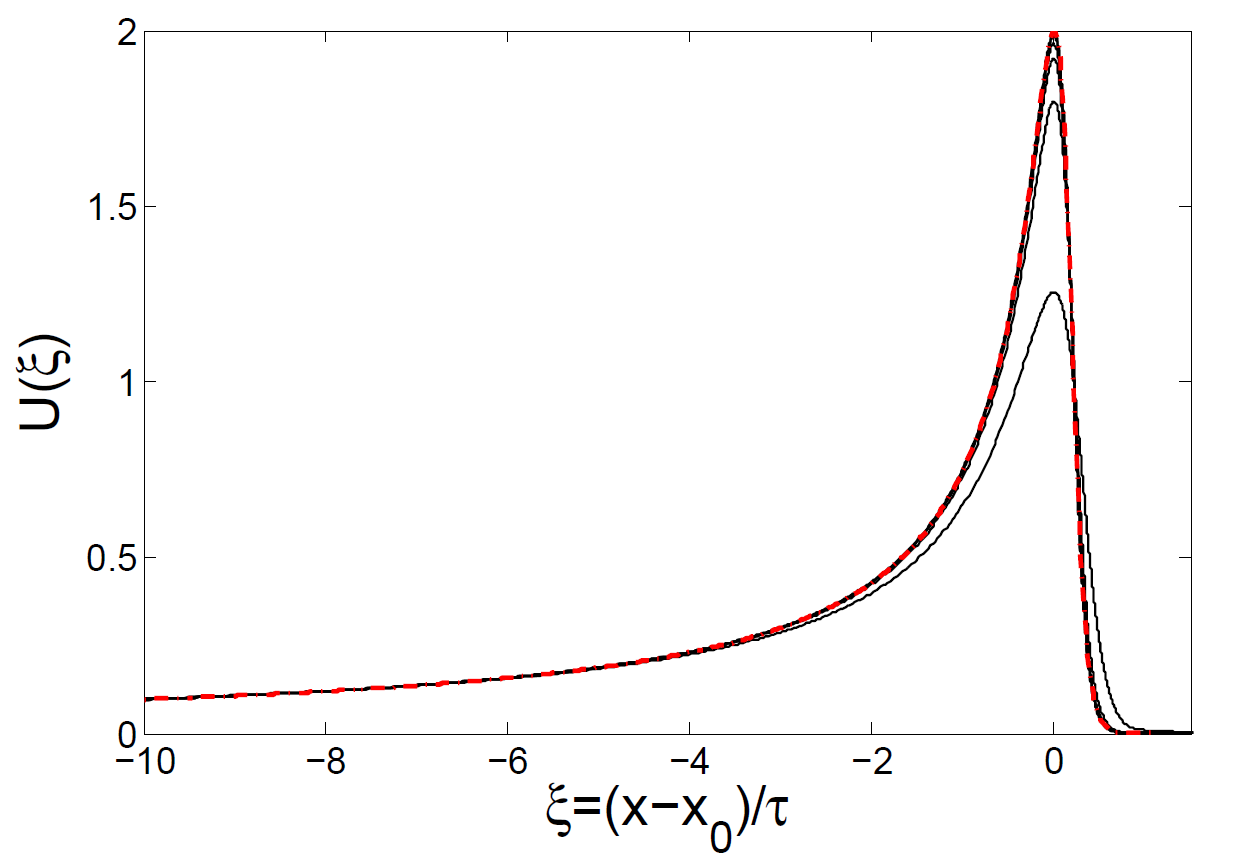}
\includegraphics*[width=0.47\textwidth]{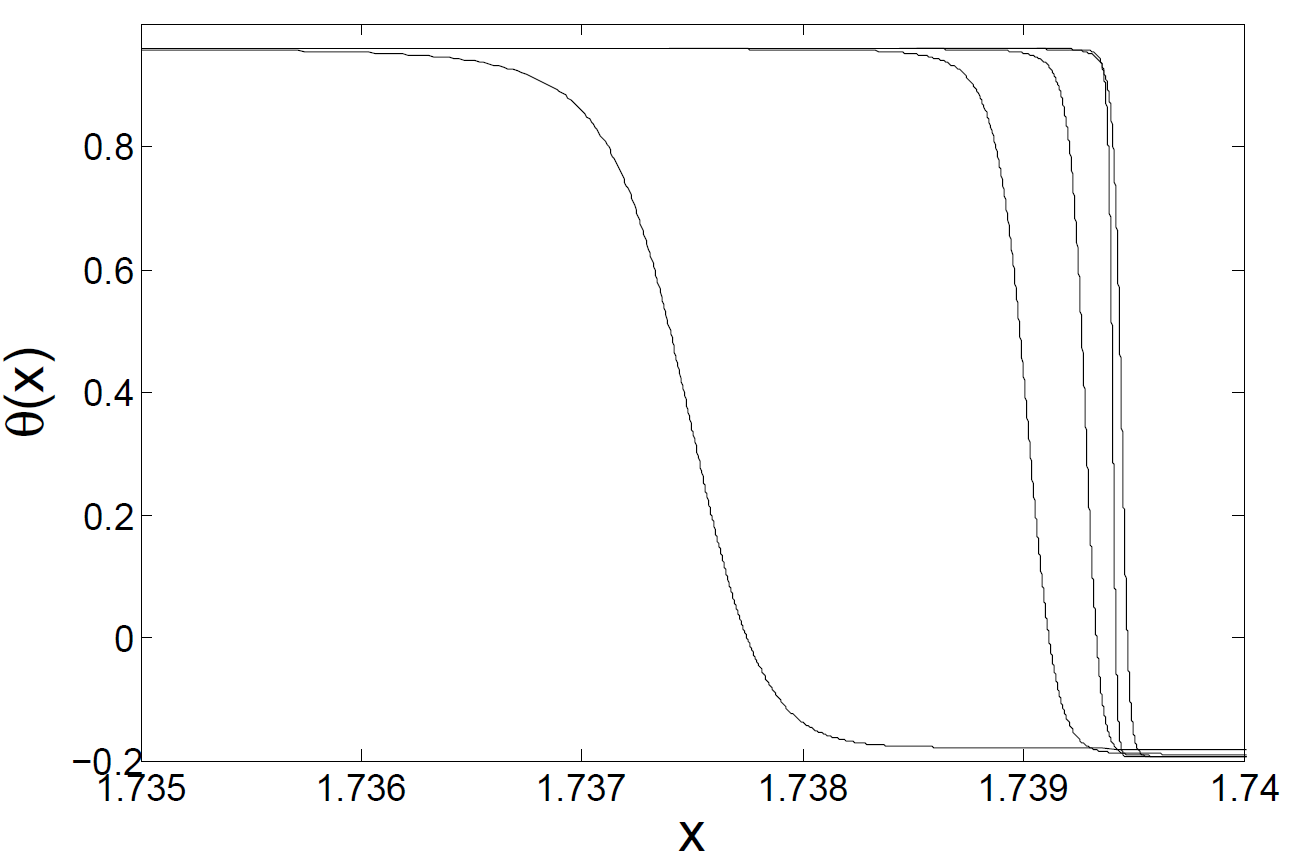}
\includegraphics*[width=0.47\textwidth]{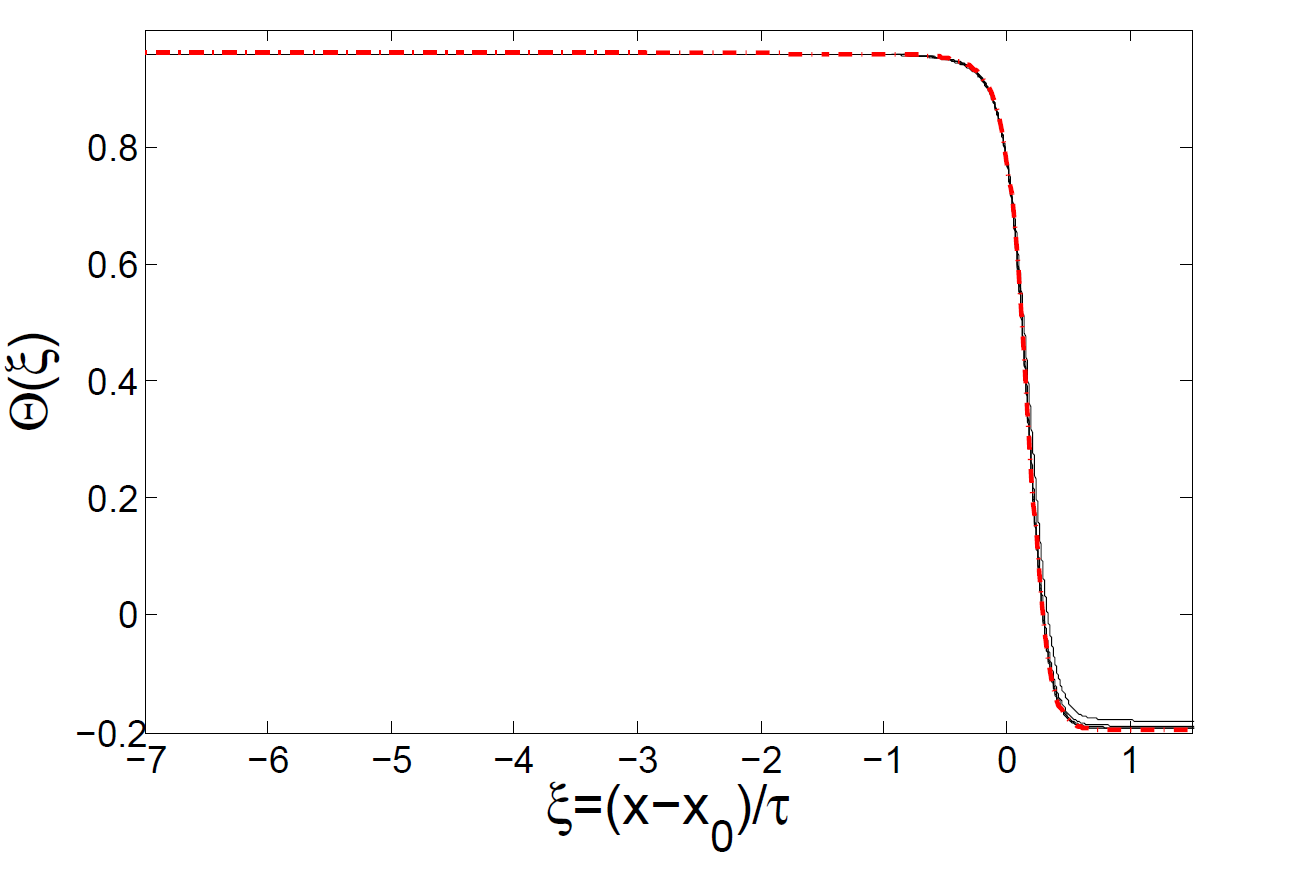}
\caption{Similarity description of the pinch region. 
Five snapshots of the velocity (first row) and temperature
(second row) profiles taken at times
$t_1=4.032,\,t_2=5.138,\,t_3=5.987,\,t_4=6.800,\,t_5=7.6410$ (black lines).
On the left, numerical solution using the same simulation as in
Fig.~\ref{pinch_ex}. On the right, profiles have been rescaled according
to \rf{SSc}, with exact solutions \rf{V_U_2} and \rf{pinch_theta}
(red dash-dotted lines) superimposed. } 
\label{pinch_ss}
\end{figure}

Differentiating \eqref{V_U_2} with respect to $\xi$ yields
\begin{align*}
\frac{w U_p^2}{4\sqrt{\cal P}^{-1}{\cal O}} =\left\{\begin{array}{ll}
-C_+ \frac{dw}{d\xi}
w\left(\sqrt{\cal P}^{-1}-w\right)^{\frac{2-4{\cal P}}{4{\cal P}-1}}
\left(4\sqrt{\cal P}-w\right)^{\frac{1-8{\cal P}}{4{\cal P}-1}} &\mbox{for } {\cal P}\not=\frac{1}{4},\\
-C_+ \frac{dw}{d\xi}\frac{w}{(2-w)^3}\exp\left\{\frac{2}{w-2}\right\}&\mbox{for } {\cal P}=\frac{1}{4},
\end{array}
\right.
\end{align*}
which, upon substituting \eqref{V_U_2} back in, can be integrated to give
\begin{align}
\xi=\left\{ 
\begin{array}{ll}
-\frac{4{\cal O}\sqrt{\cal P}^{-1}}{C_+}
\int_0^w\left(4\sqrt{\cal P}-s\right)^{\frac{1}{4{\cal P}-1}}\Big/
\left(\sqrt{\cal P}^{-1}-s\right)^{\frac{4{\cal P}}{4{\cal P}-1}} ds  &\mbox{for } {\cal P}\not=\frac{1}{4},\\
-\frac{8{\cal O}}{C_+}
\int_0^w \exp\left\{-\frac{2}{s-2}\right\}/(2-s) ds  &\mbox{for } {\cal P}=\frac{1}{4}.
\end{array}
\right.
\lb{V_U_3}
\end{align}
In \rf{V_U_3}, the origin $\xi=0$ has been chosen arbitrarily as the point
with $U_p'(0)=0$ and $U_p(0)=U_{max}$, with the maximum given by
\begin{align}
U_{max}=\left\{\begin{array}{ll}
C_+/4^\frac{4{\cal P}}{4{\cal P}-1} &\mbox{for } {\cal P}\not=\frac{1}{4},\\
C_+/(2e)&\mbox{for } {\cal P}=\frac{1}{4}.
\end{array}
\right.
\lb{U_max}
\end{align}
Combining \eqref{V_U_2} with \eqref{V_U_3} yields a parametric
representation of the pinch profile $U_p(\xi)$ with respect to the parameter
$w\in(-\infty,\,\sqrt{{\cal P}}^{-1})$ (for a typical profile see
Fig.~\ref{pinch_ss}). For $w\rightarrow-\infty$, \eqref{V_U_3} implies that
\[
\xi \approx \frac{4{\cal O}\sqrt{\cal P}^{-1}}{C_+} \ln|w|,
\quad\forall\ {\cal P}>0
\]
and since according to \rf{V_U_2} $U_p\approx C_+/|w|$ in the same limit,
we have
\beq
H_p \approx \frac{j_0}{C_+}e^{\frac{C_+\sqrt{\cal P}}{4{\cal O}}\xi}, \quad
U_P \approx C_+e^{-\frac{C_+\sqrt{\cal P}}{4{\cal O}}\xi},\quad\forall\ {\cal P}>0.
\lb{pinch_asymp+}
\eeq
Note that we can write \rf{pinch_asymp+} in the original variables as
\beqa
u(x,t)&\approx&\frac{1}{\tau(t)}
C_+\exp\Big\{-\frac{C_+\sqrt{\cal P}}{4{\cal O}}
\frac{(x-x_0)}{\tau(t)}\Big\}\nonumber\\
&=&C_+\exp\Big\{-\frac{(C_+\sqrt{\cal P}/4{\cal O})(x-x_0)+
  \tau(t)\log[\tau(t)]}{\tau(t)}\Big\},
\lb{Exp_As}
\eeqa
a representation which will turn out to be useful in the next  subsection
for matching to the solutions in the transitional region (cf. \rf{U_tr}).

The temperature profile is found from \eqref{h_theta} and
\rf{V_U_2} to be
\begin{align}
\Theta_p =\left\{ 
\begin{array}{ll}
\theta_l - \frac{j_0 C_+}{\cal M}\sqrt{\cal P}
\left(\frac{\sqrt{\cal P}^{-1}-w}
{4\sqrt{\cal P}-w}\right)^{\frac{4{\cal P}}{4{\cal P}-1}}  &\mbox{for }
{\cal P}\not=\frac{1}{4},\\
\theta_l - \frac{j_0 C_+}{2\cal M}\exp\left\{\frac{2}{w-2}\right\}
&\mbox{for } {\cal P}=\frac{1}{4}.
\end{array}
\right.
\lb{pinch_theta}
\end{align}
In Fig.~\ref{pinch_ss}, the similarity description \rf{SSc}
of the pinch region is tested against a typical numerical simulation
of the original system \rf{SSM}. On the left, we show the raw data
close to the pinch point $x_0$, while rescaled profiles are shown on the
right. Once sees very good convergence toward the exact solutions \rf{V_U_2}
and \rf{pinch_theta}, which are shown as the red dash-dotted lines. 

By taking the limit $w\rightarrow -\infty$, which corresponds to
$\xi\rightarrow\infty$, we find the following condition on the jump
of the temperature across the pinch-off (see Fig. \ref{overview}):
\beq
\theta_r = \theta_l - \frac{j_0 C_+}{\cal M}\sqrt{\cal P} < \theta_l,
\quad\forall\ {\cal P}>0;
\lb{theta_r}
\eeq
in particular, it shows that necessarily $\theta_r<\theta_l$.
It is thus seen from
\eqref{pinch_asymp+} that $H_p$ grows exponentially, which does not
match the drop profile, which has a finite slope \eqref{ConAng}. 
This means we need another region between the pinch region II and
the drop IV, which we call as the transition region.

\subsection{III: transition region}

Here we use the same similarity form \eqref{SSc} as before, but the balance
is different. On account of flux conservation, we have
$\alpha_1 + \alpha_2 = 1$ as before. We also require the transitional
solution to match onto the linear drop profile for $\xi\rightarrow \infty$,
which implies that $\alpha_1 = \beta$. Finally, we expect surface tension
to enter the force balance \eqref{SSM1}, so that from a balance between
surface tension and viscous forces we have
$\alpha_1-3\beta = \alpha_2 - 2\beta$. From these conditions we deduce
the exponents $\alpha_1 = 1$, $\alpha_2 = 0$, and $\beta = 1$, and
the similarity solution becomes
\beq
h(x,t)=\tau H_t\left(\xi\right),\quad
u(x,t)=U_t\left(\xi\right),\quad
\theta(x,t)= \Theta_t\left(\xi\right), \quad
\xi = \frac{x-x_1(t)}{\tau},
\lb{SSc_trans}
\eeq
where $x_1(t)$ denotes the center of the transition region, which will
be shown below to be slightly different from the pinch point $x_0$.
The similarity equations corresponding to \rf{SSc_trans} are 
\begin{subequations}
\label{SSMr3}
\begin{align}
j_0&=H_tU_t
\label{SSM2r2}\\
0&=H_t''' - 4{\cal O}\frac{(U_t'H_t)'}{H_t}
\label{SSM1r2}\\
0&=\left(H_t\Theta_t'\right)'.
\label{SSM3r2}
\end{align}
\end{subequations}
Here in the force balance \eqref{SSM1r2}, only {\it surface tension and
viscosity} come in at leading order $O(\tau^{-2})$.

Once more, we insert $U_t' = -j_0H_t'/H_t^2$ into \eqref{SSM1r2}
and integrate once, to obtain
\beq
C_3 = H_t''H_t -\frac{H_t'^2}{2} - 4{\cal O}j_0\frac{H_t'}{H_t}.
\lb{trans_int}
\eeq
For \eqref{SSM3r2} to be consistent with a general $H_t$-profile,
$\Theta = \theta_r$ must be a constant, which is consistent with
\eqref{theta_r}, where we matched the temperature profile in the
pinch region directly to the constant value $\theta_r$. 
In order to match to the constant slope \eqref{ConAng} for
$\xi\rightarrow\infty$, we must have $H_t\approx s\xi$ for
$\xi\rightarrow\infty$, and thus $C_3 = -s^2/2$. 

Rescaling \eqref{trans_int} according to
\beq
H_t = \frac{8{\cal O}j_0}{\sqrt{2}s}f\left(\zeta\right), \quad
\zeta = \frac{s^2}{8{\cal O}j_0}\xi,
\lb{trans_resc}
\eeq
it turns into
\beq
-1 = f''f - f'^2/2 - f'/f.
\lb{trans_int_scal}
\eeq
Putting $p(f) = f'$, the phase plane representation of \eqref{trans_int_scal}
is 
\beq
p' = \frac{p}{2f} + \frac{1}{f^2} - \frac{1}{pf}. 
\lb{trans_int_phase}
\eeq
In order to match to \eqref{pinch_asymp+}, $H_t$ must behave exponentially
for $\xi\rightarrow-\infty$, which means that $p \sim f$ near the origin
of the $p-f$ plane. On the other hand, for $\xi\rightarrow\infty$
we have seen that $H_t\approx s\xi$, and so $p\approx \sqrt{2}$
for $f\rightarrow\infty$. 

\begin{figure}
\centering
\includegraphics[width=0.6\textwidth]{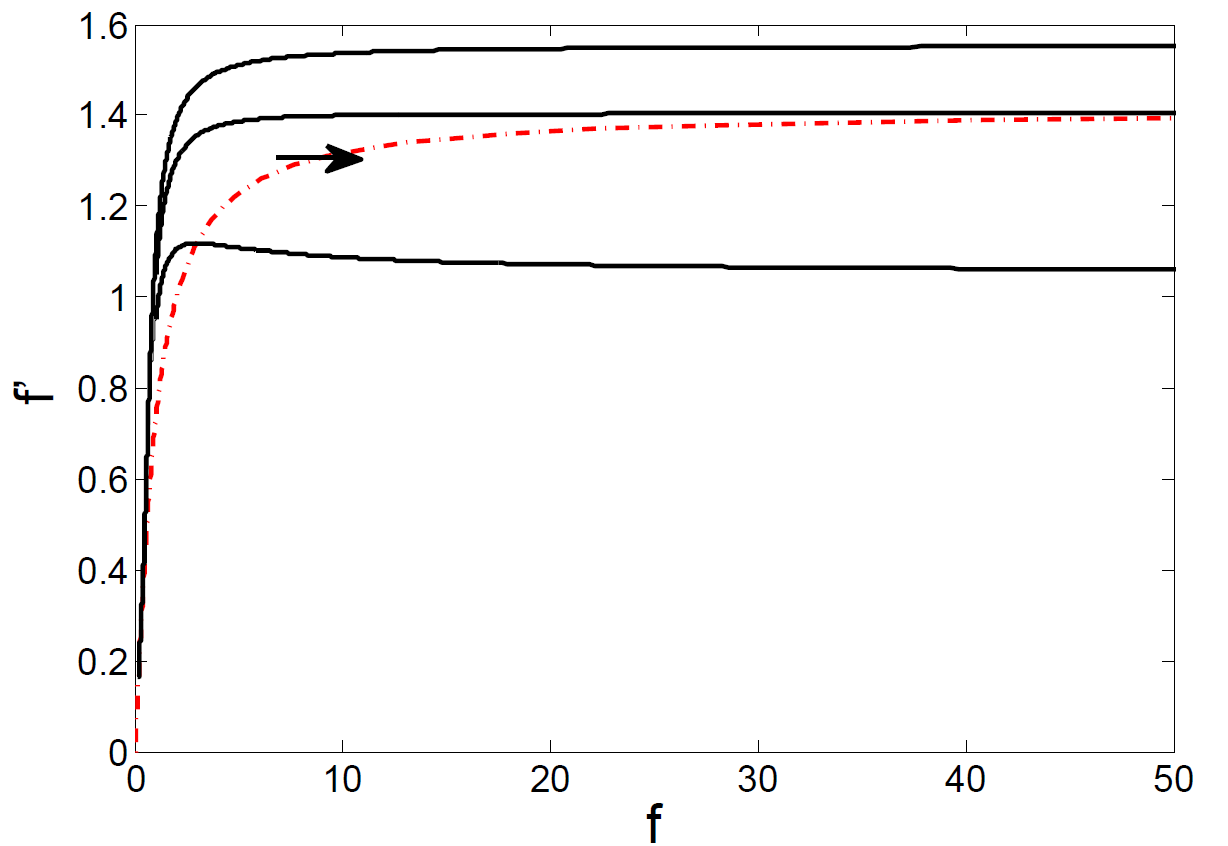}
\caption{Phase-plane portrait of the ODE \eqref{trans_int_phase}.
The solid lines correspond to the orbits and the red dash-dotted line is the
nullcline $p=0$.}
\label{trans_phase}
\end{figure}
The nullcline $p'=0$ is the curve
\beq
f = \frac{p}{1-p^2/2}, 
\lb{null}
\eeq
which is shown in the phase portrait, Fig.~\ref{trans_phase},
together with some typical solutions of \eqref{trans_int_phase}. 
An inspection of the phase plane reveals that that there is a unique
orbit that approaches the nullcline asymptotically as $f\rightarrow\infty$. As
seen from \eqref{null}, this is the solution that has the right
asymptotics for $f\rightarrow\infty$. 

For the solutions shown in Fig.~\ref{trans_phase} a more careful analysis
at the origin of the phase plane is necessary. Assuming a
regular expansion yields the series
\[
p = f + \frac{f^3}{2} + \dots \equiv p_0(f),
\]
which has no free parameters. To find the missing degree of
freedom, we put $p(f) = p_0(f) + \delta(f)$, and linearize in
$\delta$ to find
\beq
\delta' = \delta\left(\frac{2}{f}-\frac{p_0'}{p_0}+\frac{1}{p_0 f^2}\right)
\approx \frac{\delta}{f^3}, 
\lb{delta}
\eeq
for small $f$. Making the WKB ansatz
\[
\delta = \delta_0 e^{-A f^{\alpha} + \dots},
\]
a leading-order balance as $f\rightarrow\infty$ yields $\alpha = -2$
and $A = 1/2$. Thus close to the origin, we arrive at the representation
\beq
p = f + \frac{f^3}{2} + \dots + \delta_0e^{-\frac{1}{2f^2}}, 
\lb{origin}
\eeq
where the degree of freedom is in the parameter $\delta_0$. Now one can
solve \eqref{trans_int_phase} by shooting from the origin to infinity,
as shown in Fig.~\ref{trans_phase}. The value of $\delta$ is varied
until the solution asymptotes to the correct value $p = \sqrt{2}$. 
\begin{figure}[ht]
\centering
\includegraphics*[width=0.45\textwidth]{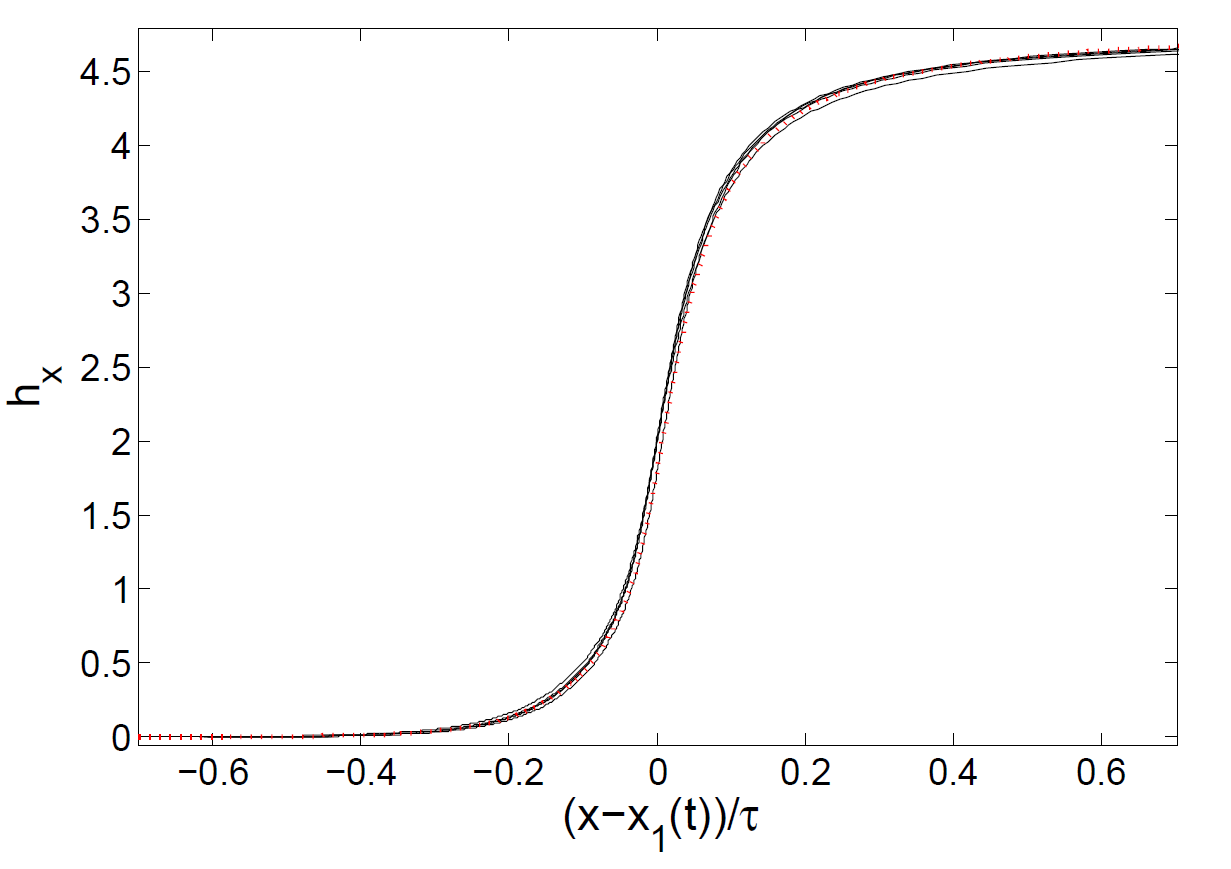}
\includegraphics*[width=0.45\textwidth]{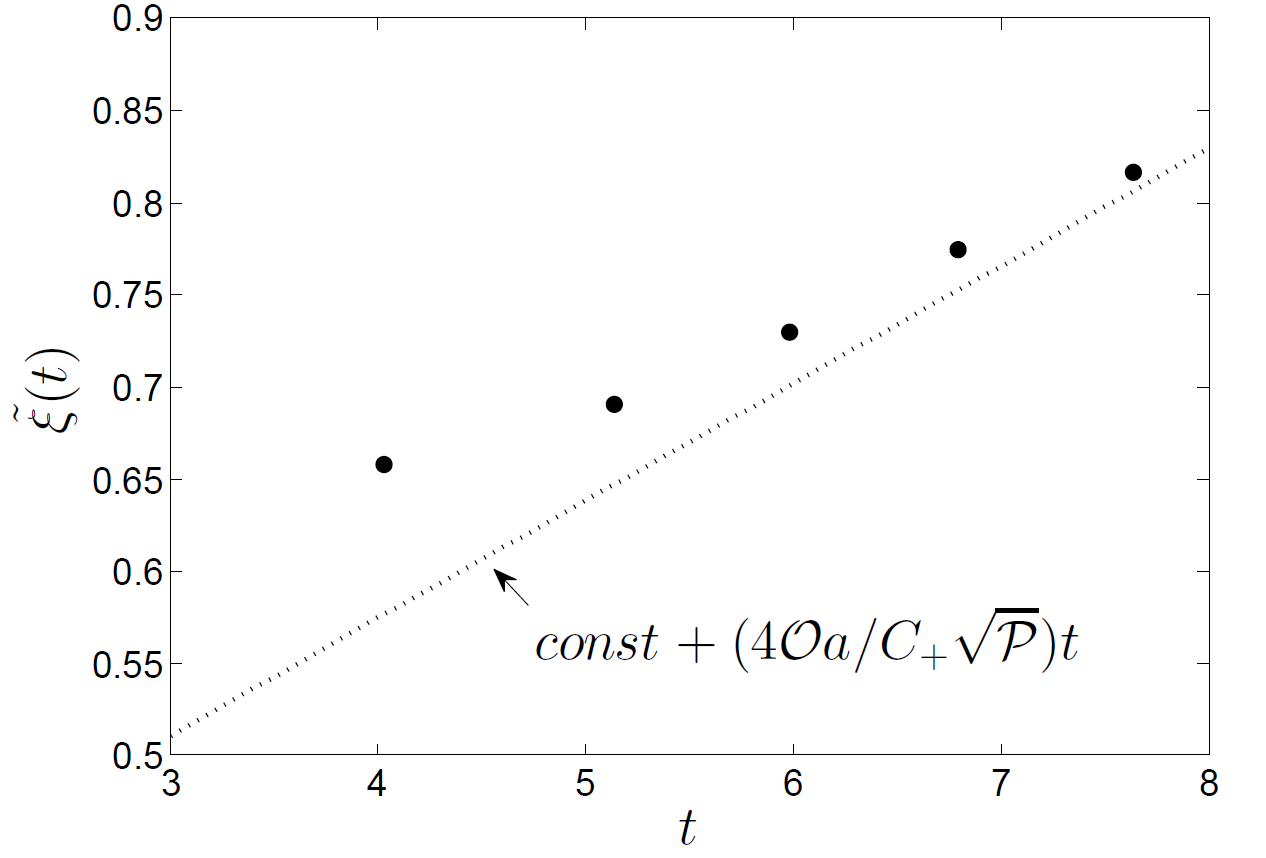}
\caption{Similarity description of the transition region. Left: 
five snapshots of the spatial derivative of the height profile
$h_x(x,t) = H_t'(\xi)$, using the 
same simulation as in Fig.~\ref{pinch_ex}, for times
$t_1=4.032,\,t_2=5.138,\,t_3=5.987,\,t_4=6.800,\,t_5=7.6410$ (black lines).
Data from the transition region is shown in self-similar variables
\rf{SSc_trans}. Each curve is shifted such that $H_t'(0)=2$, which defines 
$x_1(t)$; the collapsed profiles are compared to the solution of 
\rf{trans_int} (red dotted line), obtained by the shooting method.
Right: the corresponding values of
$\tilde{\xi}_i = (x_1(t_i) - x_0)/\tau(t_i),\ i=1,...,5$ (dots) compared to
the prediction \rf{x_1} (dotted line). The value of $a$ is taken from
Fig.~\ref{pinch_film}. }
\label{transition_ss}
\end{figure}

Once we have obtained $p(f)=f'(\zeta)$, we find $f(\zeta)$ by
(numerical) integration and thus $H_t(\xi)$ from \eqref{trans_resc}. 
The derivative $H_t'(\xi)$ is shown on the left of Fig.~\ref{transition_ss} 
as the red dotted line, and compared to $h_x$, as found from a numerical
simulation of the full system \rf{SSM}. Allowing for a horizontal shift
(which determines $x_1(t)$, see below), excellent agreement is found. 

For small $f$ ($\xi\rightarrow-\infty$), $f' = f$ yields
\[
f \approx f_0 e^{\zeta} = f_0 e^{\frac{s^2}{8{\cal O}j_0}\xi};
\]
which transforms to the asymptotics
\beq
H_t\approx B e^{\frac{s^2}{8{\cal O}j_0}\xi}\quad\text{and} \quad
U_t\approx\frac{j_0}{B}e^{-\frac{s^2}{8{\cal O}j_0}\xi}
\quad\text{as}\ \xi\go-\infty,
\lb{H_U_tr}
\eeq
as well as
\beq
u(x,t)\approx \frac{j_0}{B}\exp\left\{{-\frac{s^2}{8{\cal O}j_0}
\frac{x-x_1(t)}{\tau(t)}}\right\}
\lb{U_tr}
\eeq
in the original variables. Comparing to the asymptotics in the
pinch-off region \eqref{Exp_As}, we find the following matching conditions:
\beqa
\lb{C_+}
C_+&=&\frac{s^2}{2 j_0\sqrt{\cal P}},\\
\lb{B}
B&=&\frac{j_0}{C_+},\\
\tilde{\xi}(t)=\frac{x_1(t)-x_0}{\tau(t)}&=&\frac{\log(\tau(t))}{C_+\sqrt{\cal P}}=
\log \tau_0+\frac{4{\cal O}a}{C_+\sqrt{\cal P}}t>0,
\lb{x_1}
\eeqa
where in the last equality in \rf{x_1} we have used \rf{ODEu_0}.
On the right of Fig.~\ref{transition_ss} we test \rf{x_1}, shown as the
dotted line, against the shift $x_1(t)$, as obtained from a direct
numerical simulation (dots). For large times, the dots are seen to
approach the theoretical prediction. 

\section{Discussion and conclusions}

In this study we have derived the leading order analytical structure
of self-similar solutions describing the thermal rupture of a thin viscous
liquid sheet, and provided a consistent matching of them to the outer
solutions. The leading order solutions in the film region are given by
the velocity and height profiles \rf{film_u} and \rf{Flux}, respectively.
The film thins exponentially according to \rf{ODEu_0},  while the
macroscopic drop to its right has a parabolic profile
\rf{parabola}--\rf{CLs1}, with no flow inside. We derived explicit formulas
for the self-similar solutions \rf{V_U_2}--\rf{V_U_3} and \rf{pinch_theta}
in the pinch region, and analyzed the solution \rf{H_U_tr} in the
transitional layer. Finally, the matching conditions
\rf{ConAng}, \rf{theta_r} and \rf{C_+}--\rf{B} fix all the parameters 
of the problem in terms of $a$ and $j_0$. Since $j_0$ can be normalized
to any value by choosing the origin of the time axis, the thinning 
rate $a$ is really the only unknown parameter. We have checked numerically 
that $a$ indeed depends on the fine details of the initial data and,
therefore, can only be inferred from a more refined analysis. 

Table~\ref{tab:table1} shows the dependence of rupture parameters
upon variation of the dimensionless groups of the problem; 
the initial data are held fixed. The parameter $a$, and thus
the pinch position $x_0$ varies considerably.
In turn, table~\ref{tab:table2} shows dependence of parameters upon
variation of initial data while keeping dimensionless groups fixed.
In both tables the thinning rate $a$ was determined numerically from 
the asymptotic value of the quantity 
\[
\frac{\bar{u}^2}{4\cal O}-\frac{\bar{u}''\bar{u}}{\bar{u}'}+\bar{u}',
\]
as suggested by \rf{ODEu_0}. The parameter $\tau_0$ was fixed
(similar to Fig.~\ref{pinch_film}) so that $h_f(0)=1$, which fixes
$j_0=4{\cal O}\sqrt{a}$. The constant $C_+$ was calculated by two
alternative methods: using \rf{C_+} with the contact angle $s$ defined by
\rf{ConAng}, or from \rf{U_max}. Here the maximum velocity $U_{max}$
in the pinch region was determined from $u(x,t)$, rescaled according to
\rf{SSc}.
\begin{table}[h]
\centering
\caption{Left: Dimensionless groups and macroscopic properties 
of three numerical simulations; initial data for all simulations is as in
Fig.~\ref{pinch_ex}, with $M=\pi$ and $Q=-0.2\pi$. Right: rupture parameters
as calculated from the analytical formulas derived for them. The temperature
jump condition \rf{theta_r} and the formula for the position of the
pinch point \rf{pp} are satisfied to within an accuracy of $5\cdot 10^{-3}$.}
\label{tab:table1}
{\small\begin{tabular}{c c c c c c } 
 \hline
 ${\cal O}$ & ${\cal D}$  & ${\cal M}$ & $U_{max}$ &
 $\theta_l$ & $\theta_r=Q/M$\\
\hline
0.25 & 0.25 & 10.0 & 2.0 & 0.958 & -0.195\\ 
0.125 & 0.125 & 10.0 & 3.68 & 0.98 & -0.195 \\
0.25 & 0.25 & 20.0 & 4.121 & 0.882 & -0.195\\ 
 \hline
\end{tabular}\hspace{0.5cm}
\begin{tabular}{c c c c c } 
 \hline
 $a$ & $j_0$ & $C_+$ & $\theta_l-\theta_r$ & $x_0$\\
\hline
 0.819 & 0.905 & 12.70 & 1.149 & 1.736\\ 
 1.01 & 0.503 & 23.38 & 1.197 & 1.563 \\
 0.68 & 0.825 & 26.17  & 1.080 & 1.905\\ 
 \hline
\end{tabular}}
\end{table}
\begin{table}[h]
\centering
\caption{Left: Initial data and macroscopic properties of three
numerical simulations, with dimensionless groups fixed at
${\cal O}=1/4,\,{\cal D}=1/4,\, {\cal M}=10$.
The initial data have the same values $M=\pi$ and $Q=-0.2\pi$. 
Right: rupture parameters as calculated. Both \rf{theta_r} and 
\rf{pp} are satisfied to within an accuracy of $5\cdot 10^{-3}$.}
\label{tab:table2}
\hspace{-0.8cm}
{\small\begin{tabular}{c c c c c c } 
 \hline
 $U_{max}$ & $\theta_l$ & $\theta_r=Q/M$ & $h(x,0)$ & $u(x,0)$ & $\theta(x,0)$\\
\hline
 1.581 & 0.953 & 0.0056 & 1.0& $\pi\sin(x)$ & $\cos(x)$\\ 
 1.495 & 0.909 & 0.0051 & 1.0 & 0.0 & $\cos(x)$\\ 
 1.792 & 0.955 & -0.0949 & $1-0.2\cos(x)$ & $\pi\sin(x)$ & $\cos(x)$\\ 
 \hline
\end{tabular}\hspace{0.5cm}
\begin{tabular}{c c c c c } 
 \hline
 $a$ & $C_+$ & $j_0$ & $\theta_l-\theta_r$ & $x_0$\\
\hline
 0.887 & 10.039  & 0.942 & 0.946 & 1.668\\ 
 0.905 & 9.492 & 0.951 & 0.903 & 1.651\\ 
 0.849 & 11.379 & 0.921 & 1.048 & 1.705\\ 
 \hline
\end{tabular}}
\end{table}
\begin{figure}
\centering
\includegraphics*[width=0.35\textwidth]{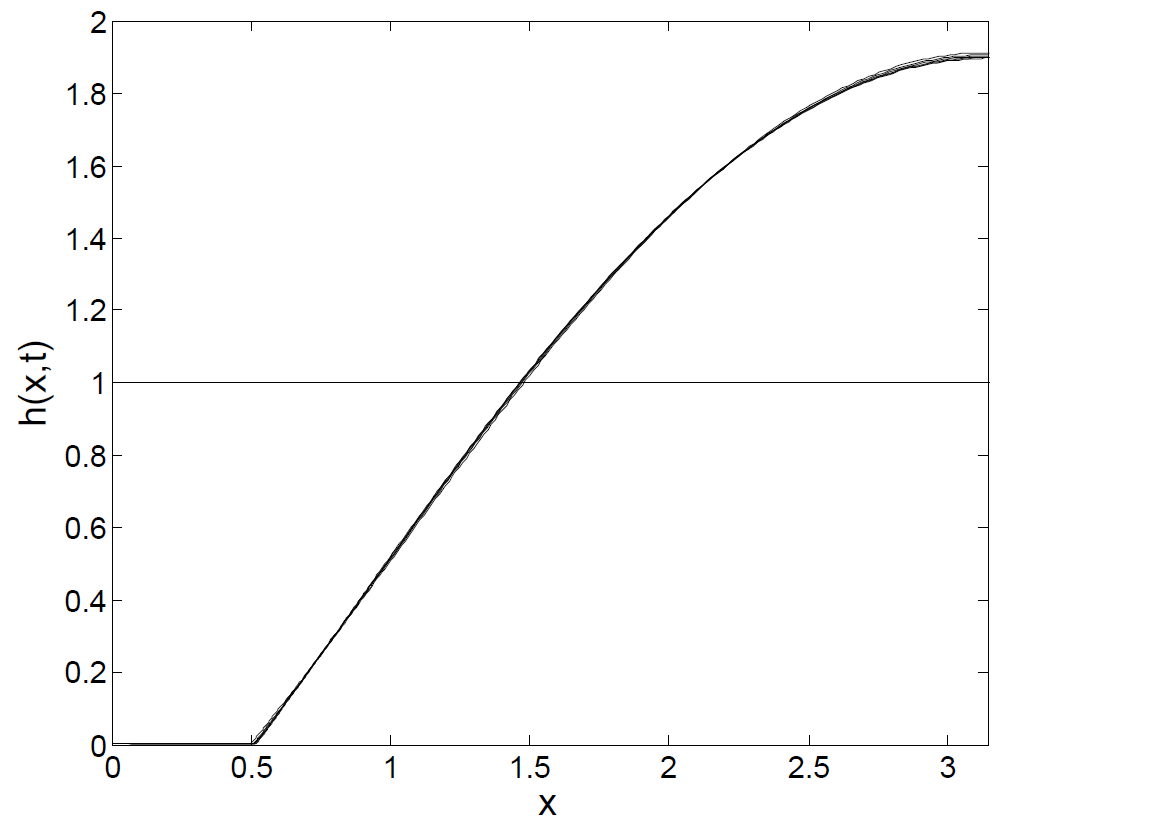}
\includegraphics*[width=0.35\textwidth]{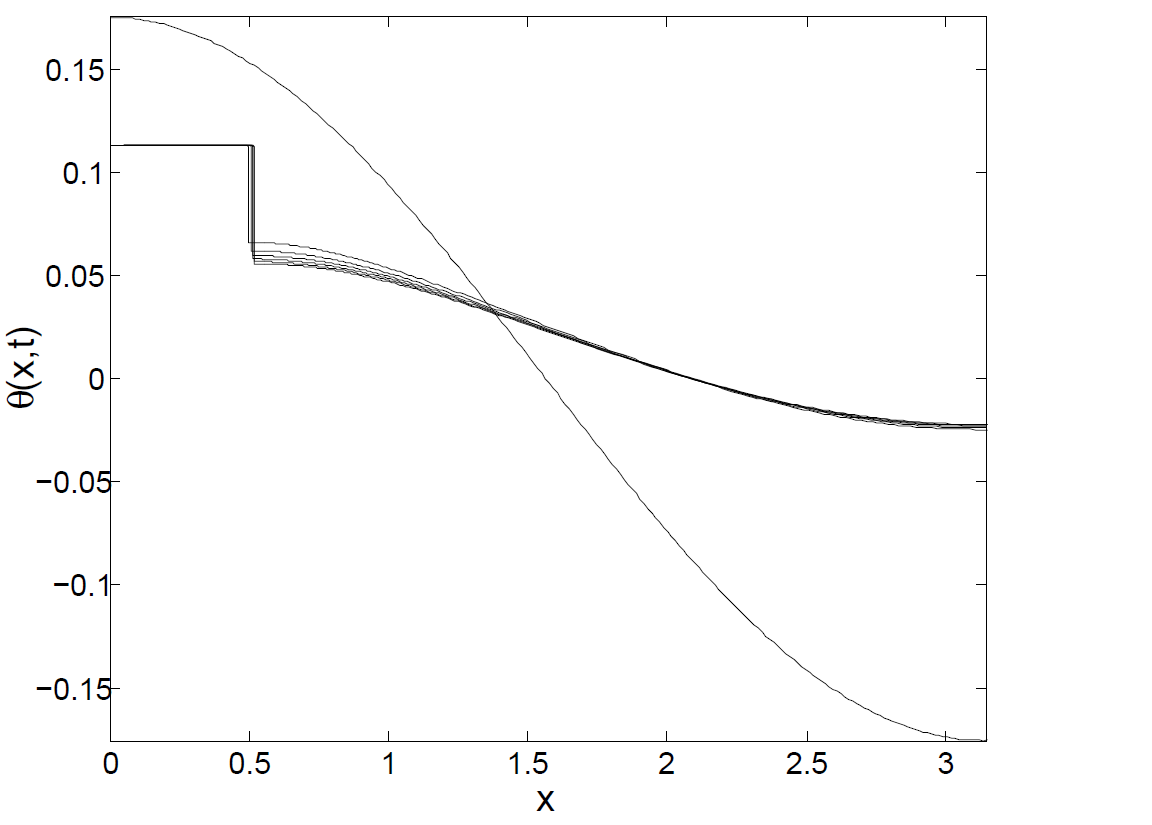}
\includegraphics*[width=0.35\textwidth]{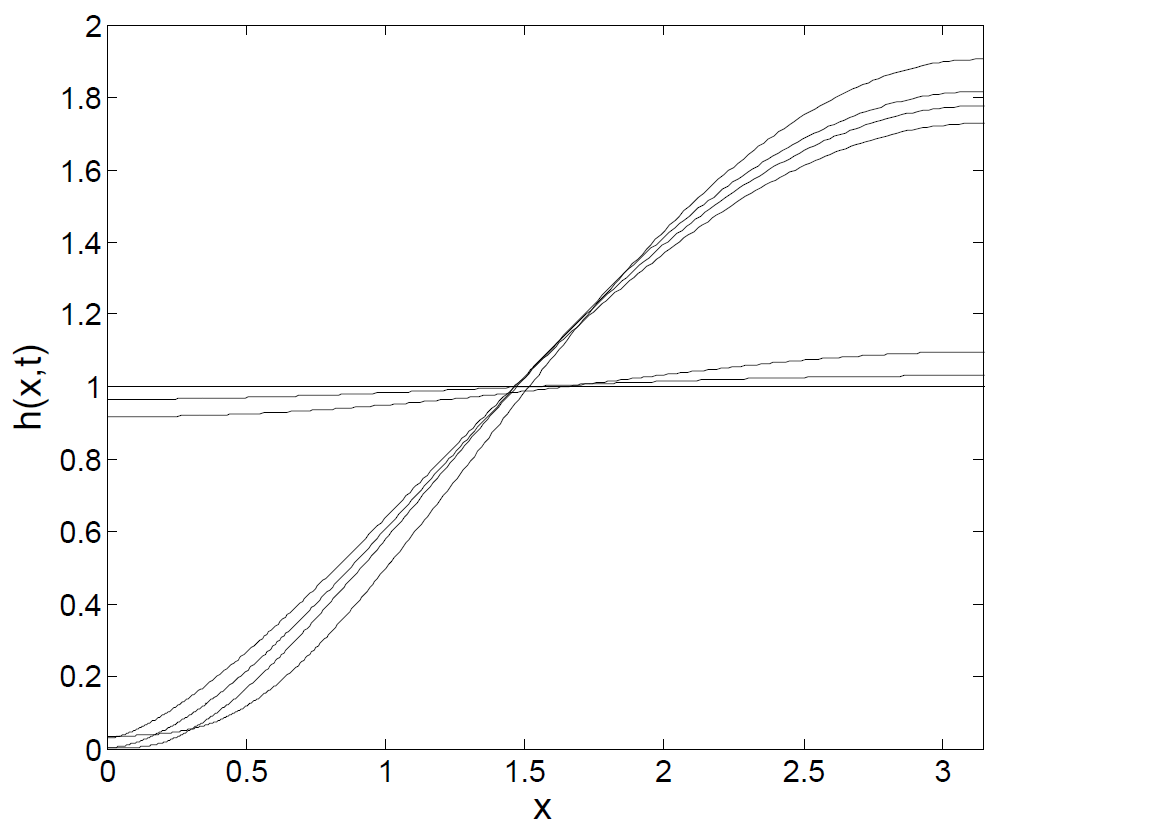}
\includegraphics*[width=0.35\textwidth]{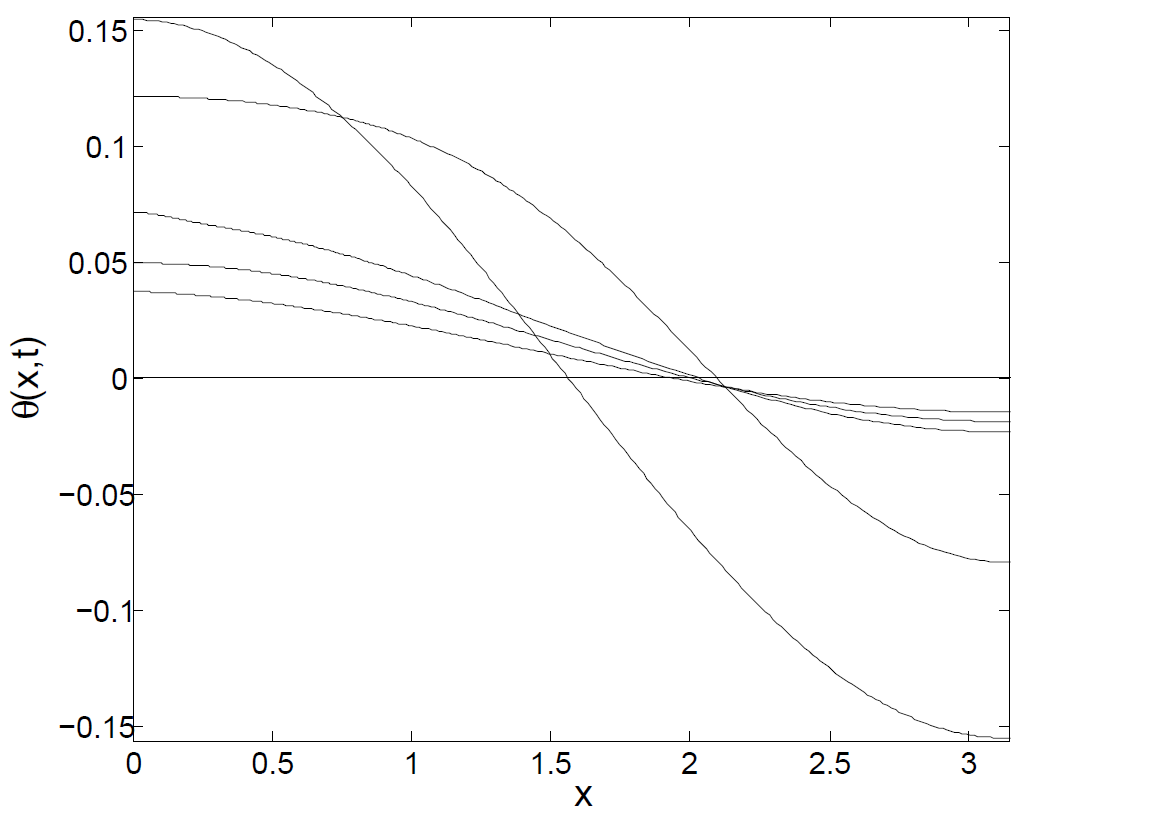}
\caption{Plots of the height (first column) and temperature (right column)
for $\Delta T=0.175 > \Delta T_{cr}$ (rupture, first row) and
$\Delta T=0.155 < \Delta T_{cr}$ (no rupture, second row). The initial data
for both simulations is given by $h(x,0)\equiv 1,\,u(x,0)\equiv 0$ and
$\theta(x,\,0)=\Delta T\cos(x)$. }
\lb{CrTemp}
\end{figure}

Since in our simulations breakup is driven by temperature gradients,
it is to be expected that there exists a critical initial temperature 
difference above which breakup occurs, while there is no breakup below
this critical value. We tested this idea using constant initial conditions
$h(x,0)\equiv 1$ and $u(x,0)\equiv 0$ for the height and velocity
profiles, respectively. The initial temperature profile  
$\theta(x,\,0)=\Delta T\sin(x+\pi/2)$ is controlled by the temperature
difference $\Delta T$. Fig.~\ref{CrTemp} confirms that for $\Delta T$
smaller than a critical value $\Delta T_{cr}$, no breakup occurs, and
instead both height and temperature relax toward constant values
(second row). If on the other hand $\Delta T > \Delta T_{cr}$, the
temperature profile develops a jump, and the height goes to zero
(first row). More detailed numerical simulations indicate that 
$\Delta T_{cr}\approx 0.16$. 

We expect a singular limiting behavior of solutions to occur when
approaching the threshold  $\Delta T_{cr}$ from above. The temperature plots
in the first row of Fig.~\ref{CrTemp} indicate that convergence toward 
the self-similar solution happens more slowly as $\Delta T_{cr}$ is
approached, especially inside of the droplet core. Moreover, the amplitude
$\theta_l-\theta_r$ of the temperature jump, and the width of the film
$(0,x_0)$ decrease to zero as $\Delta T_{cr}\go\Delta T+0$. This suggests
that the nature of the self-similar rupture changes at the critical
threshold $\Delta T_{cr}$ and the rupture, if it still occurs, should
happen then at the boundary of the interval $x=0$.

\begin{figure}[ht]
\centering
\includegraphics*[width=0.45\textwidth]{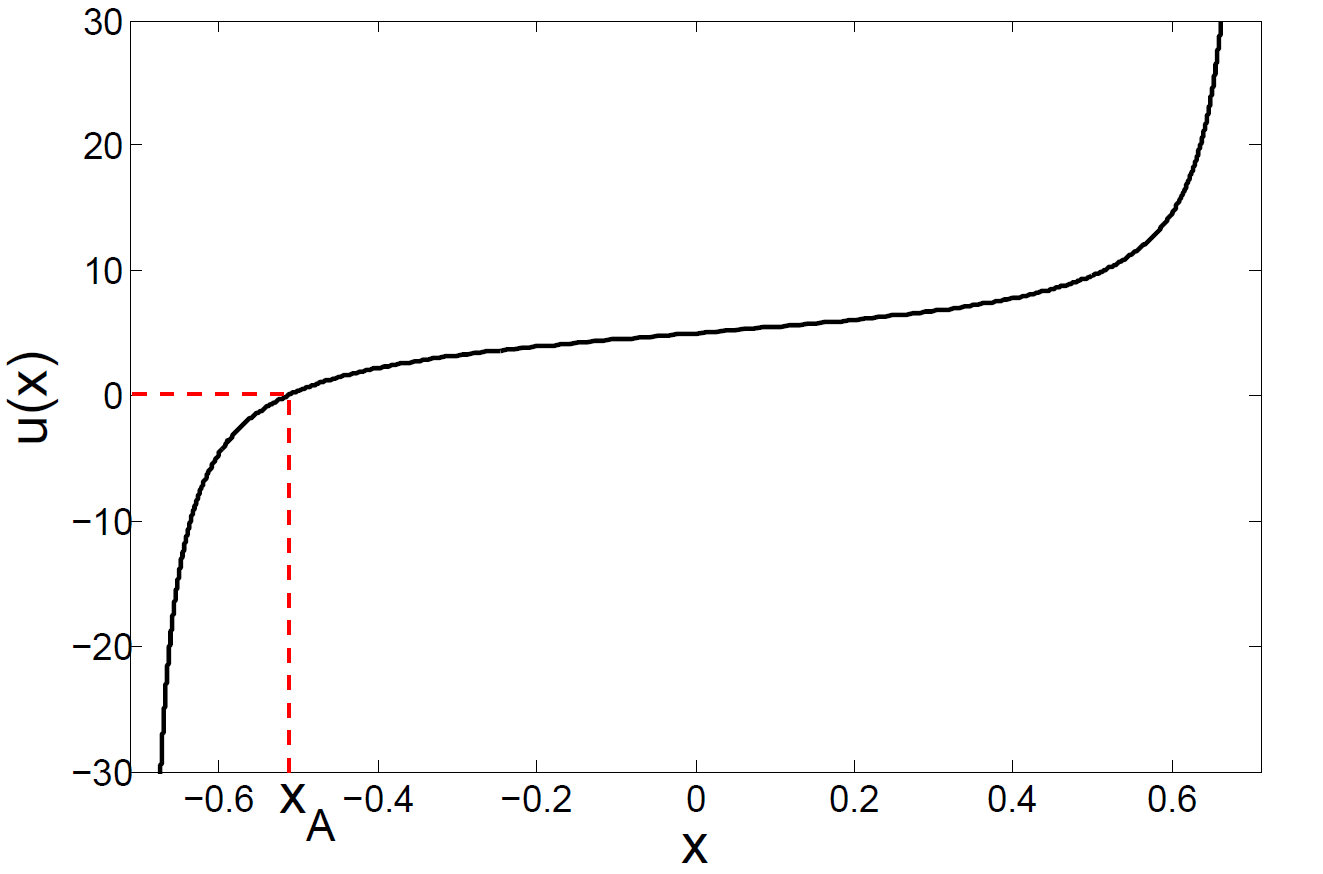}
\hspace{0.9cm}\includegraphics*[width=0.45\textwidth]{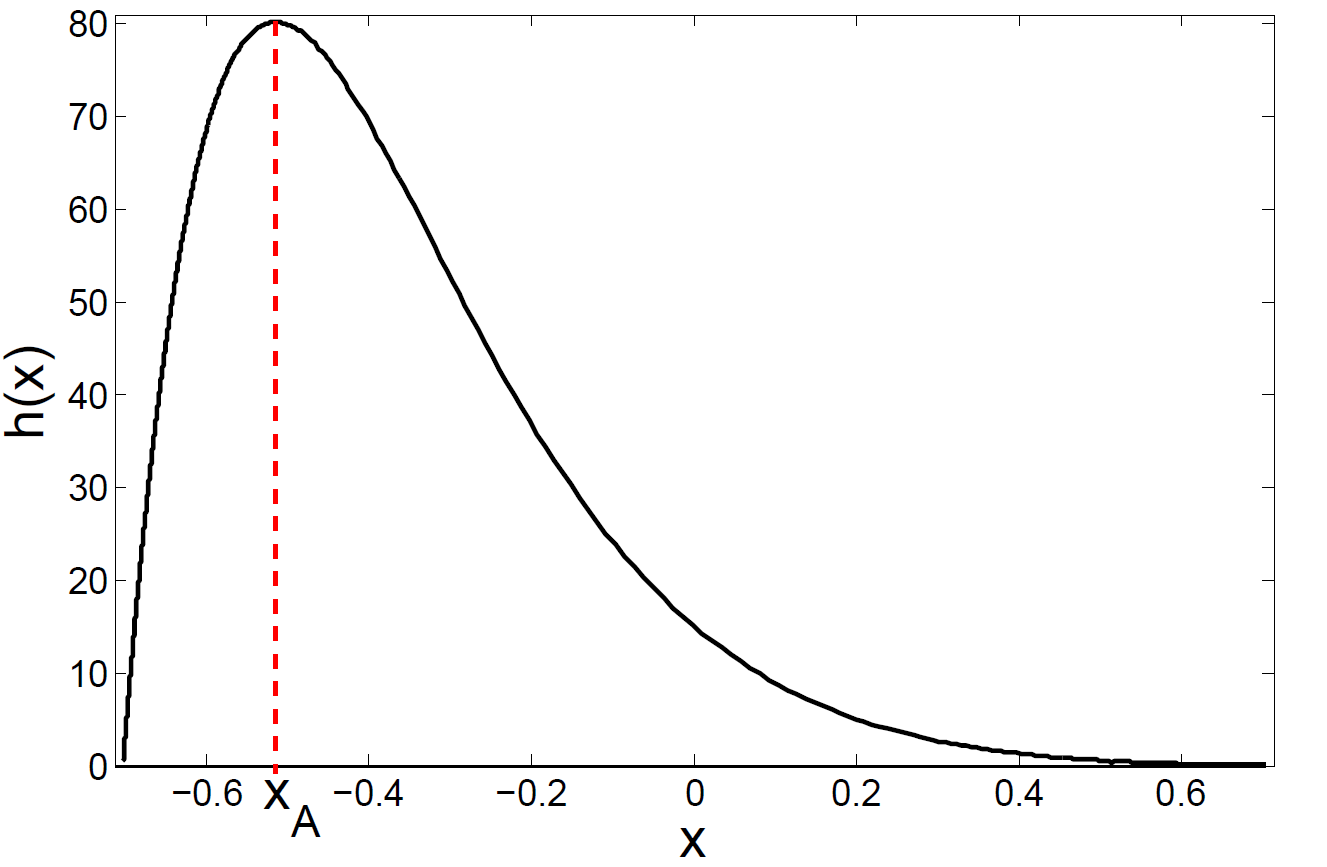}

\caption{Velocity (left) and the corresponding height profiles (right)
for the solutions to \rf{ODEu_0a} in region I, defined
in Fig.~\ref{film_phase_plot}. The velocity profile was calculated using
\rf{x_sol} for $a=30,\,A=5$. The corresponding height profiles was
calculated using \rf{flux_film} for the flux $j_f=h_fu_f$. The position
of the point $x_A$ is specified uniquely by conditions \rf{x_A}, and is shown
by the red dashed lines.}
\label{s_type}
\end{figure}

Finally, in Appendix~\ref{sec:analysis} we classify all solutions to the
ODE \rf{ODEu_0b}, which describes the velocity in the film region. 
In subsection~\ref{sub:thin} only the special solution of
type $\mathrm{I}$ (according to the classification of
Appendix~\ref{sec:analysis} and Fig. \ref{film_phase_plot}) with
$A=0$ was considered. However, our numerics indicate that for suitably
chosen initial conditions (with the same boundary conditions \rf{bc}),
solutions of type
$\mathrm{I}$ with nonzero $A$ (an example of which is shown in
Fig.~\ref{s_type}) can be realized in the thin film region. 
Namely, this happens if one evolves solutions to \rf{SSM} 
from a height profile given by a semicircle, whose maximum is located
either at $x=0$ or $x=\pi$. This drop is connected to a film region given
by the parts of the height profile shown in Fig.~\ref{s_type}, taken in the
intervals $x\in[-\pi/(2\sqrt{a-A^2}),x_A]$ or $x\in[x_A,\pi/(2\sqrt{a-A^2})]$,
respectively. Here the point $x_A$ is defined uniquely by the conditions
\beq
u(x_A)=h'(x_A)=0,
\lb{x_A}
\eeq
and is chosen to be compatible with the boundary conditions \rf{bc}.
Correspondingly, the leading order velocity in the film region is prescribed
by the corresponding parts of the type $\mathrm{I}$ solution to \rf{ODEu_0a}
with $|A|<\sqrt{a}$.

We conjecture that solution of types $\mathrm{II}-\mathrm{IV}$, represented
in the phase portrait of Fig.~\ref{film_phase_plot}, may also be realized
in more complicated rupture scenarios, for example
in the case of several pinch points separating macroscopic drops of different
sizes, which interact by virtue of small fluxes through the film regions.
This would be similar to systems considered recently by~\cite{CEFLM06,GORS08}
and ~\cite{Ki14}.  

\section*{Acknowledgments}
GK would like to acknowledge support from Leverhulme grant RPG-2014-226. 
JE and GK gratefully acknowledge the hospitality of ICERM at Brown
University, where part of this research was performed during their
participation at the trimester program "Singularities and waves in
incompressible fluids". GK gratefully acknowledges the hospitality 
of ICMAT during a research visit to Madrid.
\begin{appendix}
\section{Analysis of the velocity equation in the film region}
\label{sec:analysis}
Here we present the solution method and phase plane analysis of the
ODE \rf{ODEu_0b}:
\beq
u-\frac{a}{u}=\frac{u''}{u'}-\frac{u'}{u},
\lb{ODEu_0a}
\eeq
where for convenience we skipped overbars. We first reduce the order of the
equation by introducing a new variable $p(u)=u'(x)$. The corresponding
equation for $p(u)$ reads
\beq
u-\frac{a}{u}=p'-\frac{p}{u}.
\lb{ODEu_p}
\eeq
By introducing $\tilde{p}=p-a$, \rf{ODEu_p} reduces to the ODE
\[
u=\tilde{p}'-\frac{\tilde{p}}{u},
\]
which is invariant under the scaling $u\go Cu$ and $p\go C^2p$. Therefore,
similar to our treatment of \rf{V_U}, one can apply the substitution
$\tilde{p}=w(u)u^2$, which results in 
\[
1=w(u)+uw'(u). 
\]
This equation can be integrated to yield
\[
w(u)=-\frac{2A}{u}+1.
\]

This implies that the general solution to \rf{ODEu_p} can be characterized
completely by a one-parameter family of functions:
\beq
p(u)=(u-A)^2+a-A^2\quad\text{with}\ A\in(-\infty,\,\infty).
\lb{p_u_sol}
\eeq
The general solution to \rf{ODEu_0a} can then be obtained in the form
\[
x-\bar{x}=\int\frac{du}{(u-A)^2+a-A^2},
\]
which yields explicitly:
\begin{align}
x-\bar{x}=\left\{ 
\begin{array}{ll}
\frac{1}{2\sqrt{A^2-a}}\log
\left[\frac{u(x)-A-\sqrt{A^2-a}}{u(x)-A+\sqrt{A^2-a}}\right]
&\mbox{for } |A|>\sqrt{a},\\
\frac{1}{\sqrt{a-A^2}}\arctan\left[\frac{u(x)-A}{\sqrt{a-A^2}}\right]
&\mbox{for } |A|<\sqrt{a},\\
\frac{1}{A-u(x)} &\mbox{for } |A|=\sqrt{a}.
\end{array}
\right.
\lb{x_sol}
\end{align}
In particular, for solutions with $A\in(-\sqrt{a},\,\sqrt{a})$,
\rf{x_sol} yields the explicit solution \rf{gs_film}.
\begin{figure}
\centering
\includegraphics*[width=0.7\textwidth]{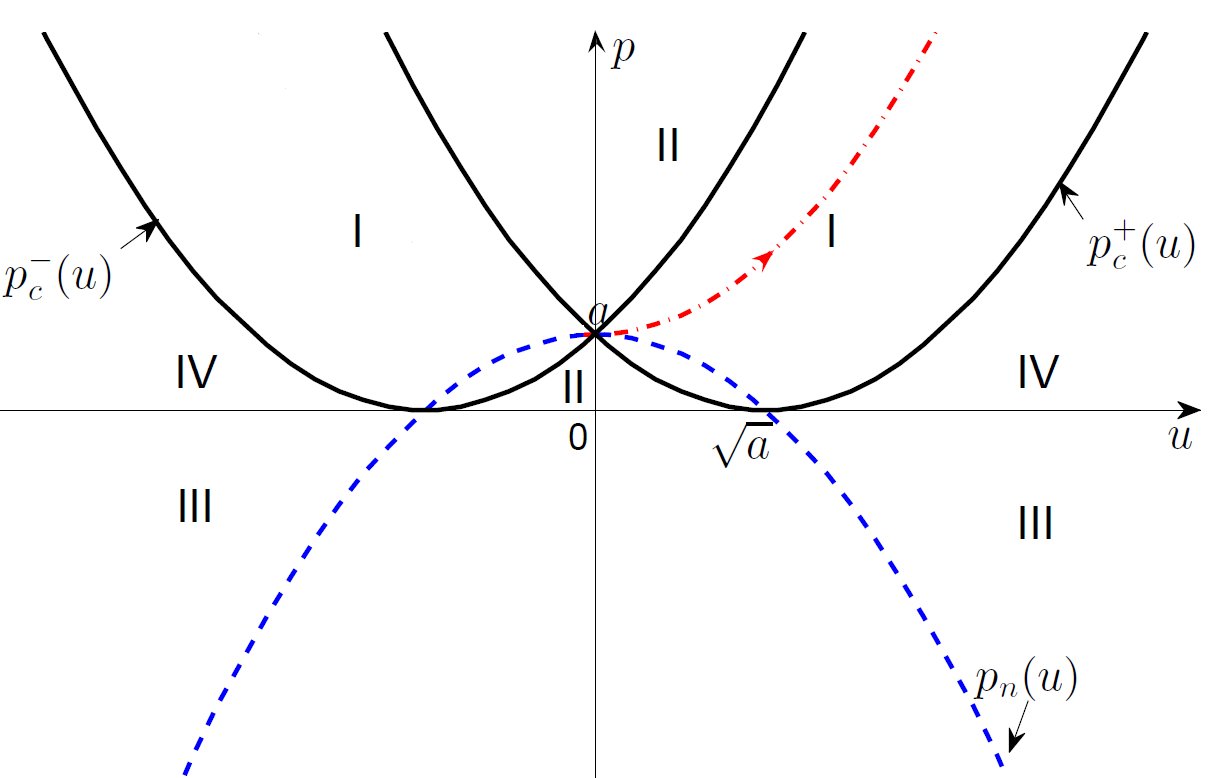}
\caption{Phase portrait for \rf{ODEu_0a}. The borders of regions
I-IV are defined by two parabolas $p_c^{\pm}(u)$
(solid black lines) and the
axis $p=0$, the nullcline is defined by the parabola $p_n(u)$
(blue dashed line). The solution curve corresponding to  \rf{film_u} is
shown as the dot-dashed red parabola.}
\label{film_phase_plot}
\end{figure}

To classify all solutions by phase-plane analysis, and to find their
regions of existence, it is useful to write \rf{ODEu_0a} as the first-order
system  
\beqa
\frac{du}{dx}&=&p,\nonumber\\
\frac{dp}{dx}&=&p\left(u-\frac{a}{u}\right)+\frac{p^2}{u} . 
\lb{pp_film}
\eeqa
Firstly, owing to the invariance $p\go p$ and $u\go -u$ of \rf{pp_film},
the phase-plane portrait is symmetric around the axis $u=0$.
Moreover, all integral curves \rf{p_u_sol} intersect at the singular point $(u=0,\,p=a)$.
The set of stationary points of \rf{pp_film} is given by the axis $p=0$,
while the nullcline $dp/dx=0$ is given by the parabola
\[
p_n(u)=a-u^2.
\]
Correspondingly, the integral curves \rf{p_u_sol} attain their minima
at the nullcline. Moreover, the axis $p=0$, together with two parabolas 
\[
p_c^\pm(u)=(u\pm\sqrt{a})^2,
\]
divide the phase plane into four regions shown as
I-IV in Fig. \ref{film_phase_plot}.
A solution to \rf{ODEu_0a} starting in one of the regions
I-IV stays in that region for all $x$.

For pinch solutions considered in this article, only those lying in region I,
characterized by $A\in(-\sqrt{a},\,\sqrt{a})$ in \rf{x_sol} and having the
explicit representation \rf{gs_film}, are relevant. By making the shift
$\bar{x}=0$ these solutions are defined in the finite interval
$x\in (-\pi/(2\sqrt{a-A^2}),\pi/(2\sqrt{a-A^2}))$. They satisfy 
\beq
u_A(\bar{x})=u_A(0)=A,
\lb{A_bc}
\eeq
and tend to infinity as $x\go \pm \pi/(2\sqrt{a-A^2})$. These two points
would correspond to pinch-off points of the full solutions to PDE system
\rf{SSM}. The special solution \rf{film_u} analyzed in
Subsection~\ref{sub:thin} corresponds to A=0, and is selected by the global
boundary conditions \rf{bc} to system \rf{SSM}, consistent with \rf{A_bc}.

Solutions lying in regions II-IV are parameterized by
constants $|A|>\sqrt{a}$. From the explicit representation \rf{x_sol} it
follows that solutions in region III are defined on the whole real
line $x\in\mR$, while solutions in regions II and IV
are defined on the half-lines $x\in (0,\,\infty)$ and $x\in (-\infty,\,0)$,
respectively.
In region III solutions are bounded and approach stationary points
$u=A\pm\sqrt{A^2-a}$ at an exponential rate as $x\go\pm\infty$. The
solutions in regions II (IV) are unbounded in the one-side limit
$x\go 0+$ ($x\go 0-$)  and approach the stationary point
$u=A-\sqrt{A^2-a}$ ($u=A+\sqrt{A^2-a}$) as $x\go\infty$ ($x\go-\infty$).
\end{appendix}

\bibliographystyle{jfm}
\bibliography{all_ref}

\end{document}